\documentclass[preprint,review,12pt]{elsarticle}

\usepackage{amssymb}
\usepackage{amsmath}
\usepackage{dcolumn}	
\usepackage{bm}			
\usepackage{color}
\usepackage{url}
\usepackage{subfigure}
\usepackage{graphicx}

\journal{Physica A}

\begin{document}
\begin{frontmatter}
\title{Constructing directed networks from multivariate time series 
       using linear modelling technique}
\author[a1]{Toshihiro Tanizawa}
\ead{tanizawa@ee.kochi-ct.ac.jp}
\author[a2]{Tomomichi Nakamura}
\ead{tomo@sim.u-hyogo.ac.jp}
\author[a3]{Fumihiko Taya}
\author[a4,a5,a6]{Michael Small}
\ead{michael.small@uwa.edu.au}
\address[a1]{Kochi National College of Technology, \\
Monobe-Otsu 200-1, Nankoku, Kochi 783-8508, Japan}
\address[a2]{Graduate School of Simulation Studies, University of Hyogo, \\
7-1-28 Minatojima-minamimachi, Chuo-ku, Kobe, Hyogo 650-0047, Japan}
\address[a3]{Singapore Institute for Neurotechnology,
Centre for Life Sciences,
National University of Singapore,
28 Medical Drive, {\#}05-COR, Singapore 117456}
\address[a4]{School of Mathematics and Statistics,
The University of Western Australia, \\
35 Stirling Hwy., Crawley, WA 6009, Australia}
\address[a5]{Complex Data Modelling Group, 
The University of Western Australia, \\
35 Stirling Hwy., Crawley, WA 6009, Australia}
\address[a6]{Mineral Resources, CSIRO, Kensington, WA 6151, Australia}
\begin{abstract}
We describe a method to construct directed networks from multivariate time series
which has several advantages over the widely accepted methods.
This method is based on an information theoretic reduction of 
linear~(auto-regressive) models.
The models are called reduced auto-regressive~(RAR) models.
The procedure of the proposed method is composed of three steps:
(i)~each time series is treated as a basic node of a network,
(ii)~multivariate RAR models are built and the constituent information 
in the models is summarized,
and (iii)~nodes are connected with a directed link based on 
that summary information.
The proposed method is demonstrated for numerical data generated by known systems, 
and applied to several actual time series of special interest.
Although the proposed method can identify connectivity, 
there are three points to keep in mind:
(1)~the proposed method cannot always identify nonlinear relationships among 
components, (2)~as constructing RAR models is NP-hard,
the network constructed by the proposed method might be near-optimal network
when we cannot perform an exhaustive search,
and (3)~it is difficult to construct appropriate networks 
when the observational noise is large.
\end{abstract}
\end{frontmatter}

\section{Introduction}
\label{sec:introduction}

Time series of natural phenomena usually show irregular fluctuations,
and it is often found that such behaviours can be attributed to time 
delays~(periodicities) in systems~\cite{Ohira-Yamane:2000}.
We consider periodicities as an important clue to understand dynamical phenomena 
in nature, irrespective of whether the data are linear or nonlinear.
When data is linear 
the information of periodicities is directly linked to essential 
understanding of the linear system generating the data.
Nonlinear data might also have periodicities.
In this case the periodicities is one of the important clues 
to understand the underlying characteristic in the data and the system.
In this paper, we propose a method to construct directed networks from 
multivariate time series from the perspective of linear periodic 
structures~(relationships).

There has been various works for constructing networks from multivariate 
time series and many applications~\cite{Kaminski-Blinowska:DTF1991,%
Kaminski-etal:DTF2001,Baccala-Sameshima:PDF2001,Mantegna:1999,Farkas-etal:2003,%
Astolfi-etal:2007,Yamasaki-etal:2008,Tsonis-Swanson:2008,Tse-etal:2010,%
Gao-Jin:2009,Nagy-etal:pigeon_flocks10,Iwayama-etal:2012,Gao-etal:network2016}.
Among them, there are two previously proposed and widely accepted approaches 
for constructing networks from multivariate time series.
One uses information from the frequency domain in which phase differences
(shifts or delays) at a frequency between two signals are 
examined~\cite{Astolfi-etal:2007}, and the other uses information from 
the time domain in which similarities between two signals are examined 
in terms of time differences~\cite{Nagy-etal:pigeon_flocks10}.

Although both  approaches have proven to be effective in various 
cases~\cite{Astolfi-etal:2007,Nagy-etal:pigeon_flocks10},
we feel that their effectiveness is constrained by the following
two possible concerns.
The first one is that, as the perspective to construct the networks
by these approaches has never been clearly specified,
what the constructed networks actually represent is not clear. 
In these approaches, multivariate auto-regressive~(MVAR) model,
the cross correlation~(CC) function, and a fixed threshold value are applied
to investigate the existence of relationship between a pair of time 
series~\cite{Astolfi-etal:2007,Nagy-etal:pigeon_flocks10}.
The second concern is that these statistical approaches
often cannot adequately capture more local, nonlinear, or non-stationary 
peculiarities of time series.
As a result, it is not clear what the constructed networks by the methods 
indicate for the data. 
Although the perspective of these approaches has never been clearly specified,
we consider that the perspective of these approaches corresponds to 
linear periodic structures, because the MVAR model and the CC function are used.
As mentioned above, both approaches are indirect methods to identify 
underlying linear periodic structures among multivariate time series.
We consider that more straightforward approach is preferable to identify 
subtle features of the structures.

In this paper we propose a method which can construct networks from 
multivariate time series reflecting their dynamical nature as faithfully as 
possible based on a firm perspective.
The proposed method utilizes a previously proposed linear model, 
the reduced auto-regressive~(RAR) model~\cite{Judd-Mees:1995,Judd-Mees:1998,%
Small-Judd:1999}.
The RAR model can precisely identify periodicities 
that are present in a time series,
irrespective of whether the data is linear or nonlinear, 
provided the time series is sufficiently long~\cite{Small-Judd:1999}.
Of course, there are restrictions when applying the proposed method. 
The RAR model cannot always identify nonlinear periodic structure in the data.
To build a RAR model we need to find the optimal subset of possible terms 
for the model, which is expected to be an NP-hard problem.
In this case, we usually use a selection algorithm,
and the obtained RAR model might be only nearly optimal.
It is also difficult to build appropriate RAR models~(and 
to construct appropriate networks) when the observational noise is large.

The paper is organized as follows.
We briefly review two widely accepted approaches as the current approaches
in Section~\ref{sec:current_approaches}.
In Section~\ref{sec:desired_networks} we identify the network we like to
construct from a given set of time series.
In Section~\ref{sec:problems_with_current_approaches}
we describe the problems with the current approaches,
and show that the current approaches cannot construct the desired networks.
In Sections~\ref{sec:new_approach} and~\ref{sec:imperfect_model_scenario} 
we introduce our method and apply the proposed method to several cases 
using simulated multivariate time series of known linear systems
where there are correct linear model systems
and the R{\"o}ssler systems where there is no correct linear system.
We discuss difficulties with building RAR models 
in Section~\ref{sec:reexamination_of_the_proposed_method}.
In Section~\ref{sec:application} we apply our method to real-world multivariate 
time series data, which are meteorological data and electroencephalography data.

\section{Current approaches}
\label{sec:current_approaches}

There are two major approaches to construct networks from multivariate 
time series, which are classified into the frequency-based approach and 
the time-based approach.
In these approaches each individual time series is treated as a basic node 
of a network and a threshold is used to test the existence of relationship 
between data.
\subsection{Frequency-based approach to network construction}
\label{sec:frequency-based_approach}

There are also two widely accepted frequency-based 
methods~\cite{Astolfi-etal:2007,Blinowska:2011,Sameshima-Baccala_book:2014}.
One is Directed Transfer Function~(DTF)~\cite{Kaminski-Blinowska:DTF1991,
Kaminski-etal:DTF2001},
and the other is Partial Directed Coherence~(PDC)~\cite{Baccala-Sameshima:PDF2001}.
DTF was proposed as a multivariate spectral measure to determine 
the directional influences between any given pair of time series 
in a multivariate dataset~\cite{Kaminski-Blinowska:DTF1991,Kaminski-etal:DTF2001}.
DTF is an estimator that simultaneously characterizes the direction and 
spectral properties of the interaction between signals.
PDC was proposed as a factorization of the Partial Coherence after DTF,
and PDC is based on MVAR coefficients transformed into the frequency 
domain~\cite{Baccala-Sameshima:PDF2001}.
Both methods are based on the MVAR model,
and the MVAR model is transformed to the frequency domain by the Fourier transform 
(or $ z $ transformation) to investigate the spectral properties.
The pair of nodes corresponding to the chosen two time series is connected 
with a directed link when a value calculated by both the methods is larger than 
an appropriately chosen threshold.
A threshold value is used to determine whether the values are large enough.
See more details on DTF and PDC elsewhere~\cite{Kaminski-Blinowska:DTF1991,%
Kaminski-etal:DTF2001,Baccala-Sameshima:PDF2001,Astolfi-etal:2007}.

Both DTF and PDC are based on the detection of phase differences 
between two signals.
Hence, we consider that both DTF and PDC make effective use of 
linear periodic structure between the data.
\subsection{Time-based approach to network construction}
\label{sec:time-based_approach}

The most extensively used method of time-based approach utilizes 
the cross correlation~(CC) function,
and a fixed threshold value is used to determine the existence of relationship
between data.
Generally, the basic procedure can be reduced to the following three steps.
\begin{enumerate}
\renewcommand{\labelenumi}{(\arabic{enumi})}

\item
  Each individual time series is considered as a basic node of a network.

\item
  To investigate the relationship among multivariate time series,
  all values of the CC function between
  the whole pairs of these time series are calculated.

\item
  The node pairs whose values of the CC function 
  are larger than an appropriately chosen threshold 
  are connected with undirected links.

\end{enumerate}
We refer to this method as the ``naive method.''
As the naive method utilizes the CC function,
we consider that the networks obtained by the naive method reflects
pairwise linear periodic structure among the data.

\subsection{How to determine the existence of relationships}
\label{sec:how_to_determine_relationships}

We need to examine the existence of relationships between two time series~(nodes)
when applying the current approaches for constructing a network.
A simple way is to use a fixed threshold value.
When the value of the CC function is larger than the threshold value
we expect that there may be some sort of relationship between the two variables,
and hence the pair are considered to be connected.
A commonly used threshold value is $ 0.5 $~\cite{Tsonis-Swanson:2008,%
Tse-etal:2010}, and we also use the value in this paper for examining
the current approaches.

Though it is possible to use a fixed threshold value for DTF and PDC, 
another approach using the surrogate data method has also been 
proposed~\cite{Kaminski-eta:DTF2001,Toppi-etal:2012}.
In this method,
the surrogate data are treated as a null case,
in which the data corresponding to a given pair of time series explicitly 
lack relationship,
and compared to those of DTF and PDC at a certain probability~$ p $.
In a sense, the value related to the surrogate data is treated as a threshold.
The surrogate data are generated as follows:
(i)~the Fourier transform~(FT) is applied to the original data, 
(ii)~randomizes the phases, 
and (iii)~then inverts the transform using the randomized 
phases~\cite{Theiler-etal:surrogate92}.
The data generated by this algorithm is often referred to as 
the FT surrogate data.
Further details of the surrogate data method and the algorithms 
are provided in Refs~\cite{Theiler-etal:surrogate92,galka_book,Michael_book}.
When the value of DTF and PDC of the original data is decided to be larger 
than that of the FT surrogate data sets with a predefined significance level,
the pair is connected.
In this paper we use the integrated value of DTF and PDC
of the original data and the FT surrogate data sets,
generate 1000~FT surrogate data sets for each pair of signals,
and the significance level is $ 5\% $~(that is, $ p < 0.05 $).

\section{Identifying networks to be constructed}
\label{sec:desired_networks}

It is important that the network constructed from multivariate time series 
is a faithful representation of the system generating the data.
In this section we describe how a network can be such a faithful
representation when the system generating the data is specified.
In the next section we will show that the current approaches cannot 
construct such a network, although there are linear periodic 
structures among multivariate time series.

When we observe time series data, the information for
the true system~(model) that generates the data is not usually available.
In this case, we take a phenomenological approach for
describing the phenomenon by building ``a model''
that reproduces the data as much as possible.
The network to be constructed should therefore represent connectivities
between the elements of such a model.
As mentioned above, we consider that the actual perspective of the current 
approaches corresponds to linear periodic structures~(relationships).
In this section we identify desired network structures reflecting this perspective
based on two artificial but possible systems.

The first system~(system~1) is described by the following 
expressions~\cite{Nakamura-etal:SSS_network2016}:
\begin{align}
 x_1(t) &= 1.3 + 0.4 \; x_1(t-1) - 0.2 \; x_1(t-3) + 0.3 \; x_2(t-4) + 0.2 \; x_4(t-7) + \varepsilon_1(t),  \label{eq:multi_linear_model1-1} \\
 x_2(t) &= 2.0 + 0.6 \; x_2(t-1) - 0.2 \; x_2(t-6) + \varepsilon_2(t),  \label{eq:multi_linear_model1-2} \\
 x_3(t) &= 2.2 + 0.2 \; x_1(t-2) + 0.3 \; x_4(t-9) + \varepsilon_3(t), \label{eq:multi_linear_model1-3} \\
 x_4(t) &= 1.3 + 0.2 \; x_1(t-2) + 0.5 \; x_4(t-1) - 0.3 \; x_4(t-3) + \varepsilon_4(t), \label{eq:multi_linear_model1-4}
\end{align}
and the second system~(system~2) is given by
\begin{align}
 x_1(t) &= 12.0 + 1.2936 \; x_1(t-1) - 0.3022 \; x_1(t-4) + 0.2019 \; x_2(t-3) + \varepsilon_1(t), \label{eq:multi_linear_model2-1} \\
 x_2(t) &= 0.3007 \; x_2(t-1) + 0.2021 \; x_2(t-6) + \varepsilon_2(t), \label{eq:multi_linear_model2-2} \\
 x_3(t) &= 1.2099 \; x_2(t-4) - 0.6023 \; x_2(t-10) + 0.9392 \; x_3(t-1) + \varepsilon_3(t), \label{eq:multi_linear_model2-3} \\
 x_4(t) &= 5.5902 \; x_2(t-3) + 0.9201 \; x_4(t-1) + \varepsilon_4(t), \label{eq:multi_linear_model2-4}
\end{align}
where $ \varepsilon_i(t)\: (i = 1, 2, 3, 4) $ are dynamic noise,
independent and identically distributed~(IID) Gaussian random variables 
with mean~zero and standard deviation~1.0 for both the systems.
That is, these systems are perturbed by dynamic noise.
The coefficients in both systems are chosen arbitrarily 
so that the generated multivariate time series do not diverge.

\begin{figure}[!t]
\centering
  \includegraphics[width=6.5cm]{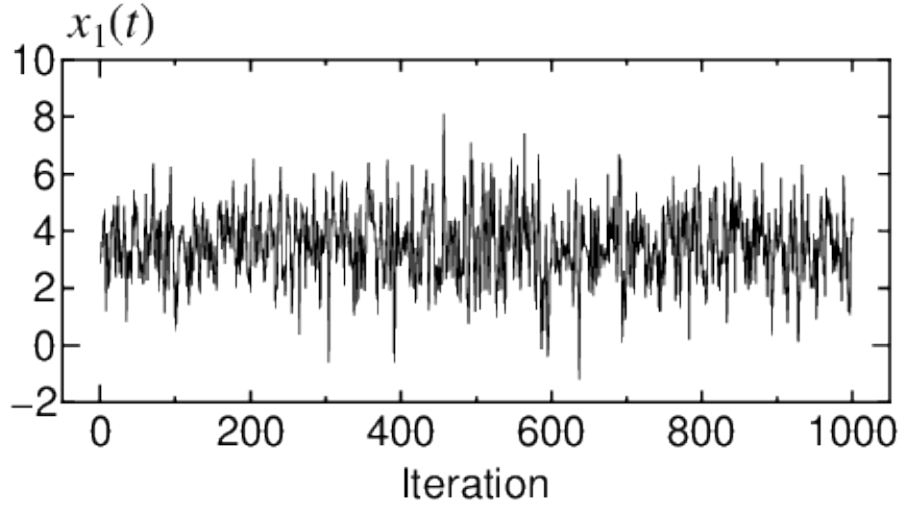}
~~\includegraphics[width=6.5cm]{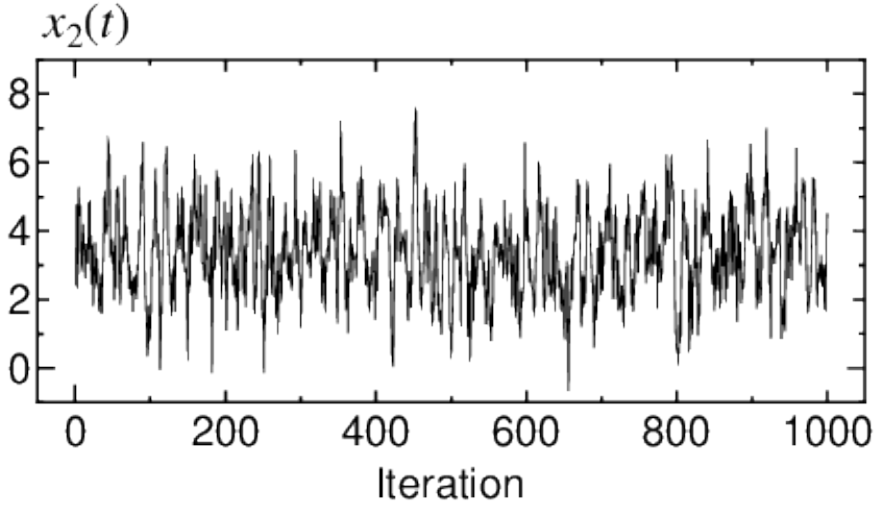}\\
  \includegraphics[width=6.5cm]{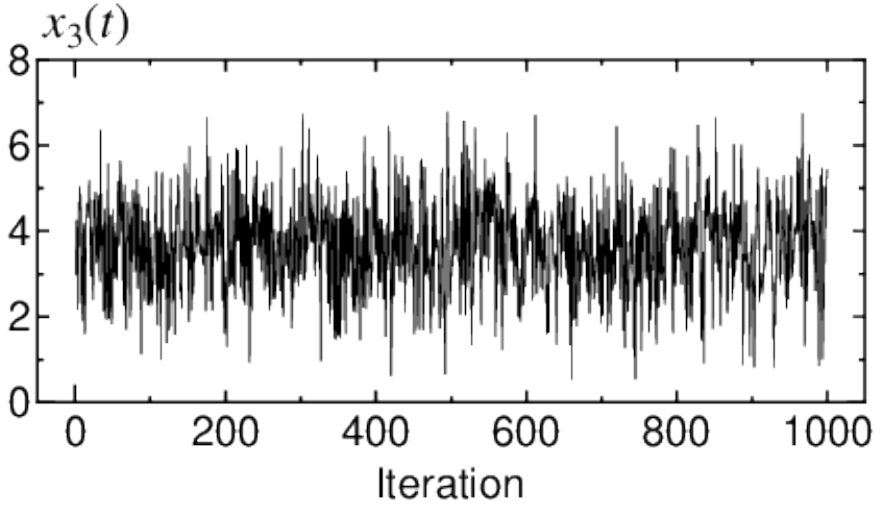}
~~\includegraphics[width=6.5cm]{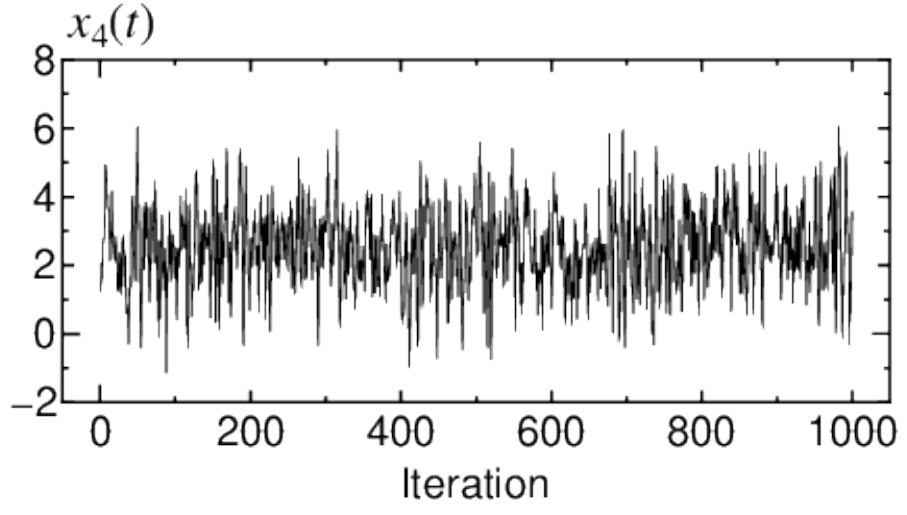}
\caption{Time series data generated by 
         Eqs.~(\ref{eq:multi_linear_model1-1})--(\ref{eq:multi_linear_model1-4}).
         Irregular fluctuations are observed in all variables 
         with similar time scales.}
\label{fig:example1_RAR}
\end{figure}

\begin{figure}[!t]
\centering
  \includegraphics[width=6.5cm]{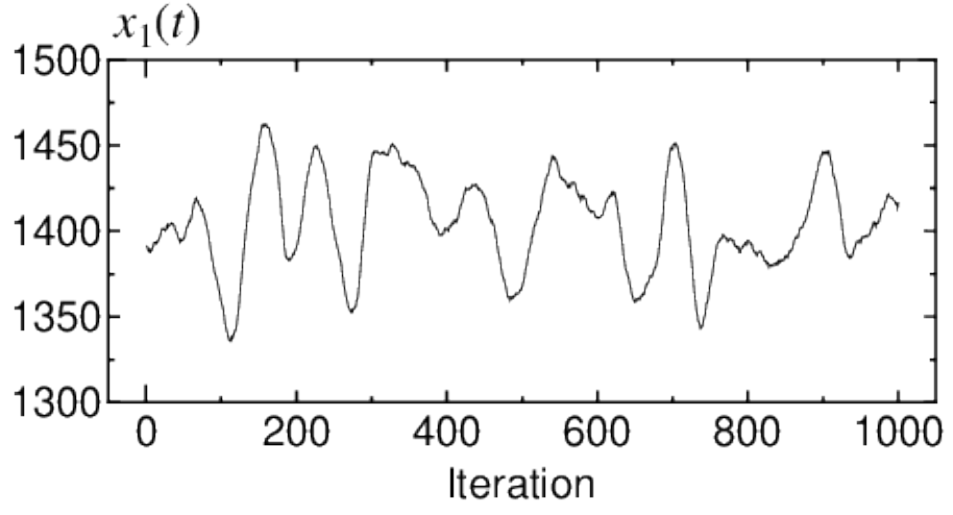}
~~\includegraphics[width=6.5cm]{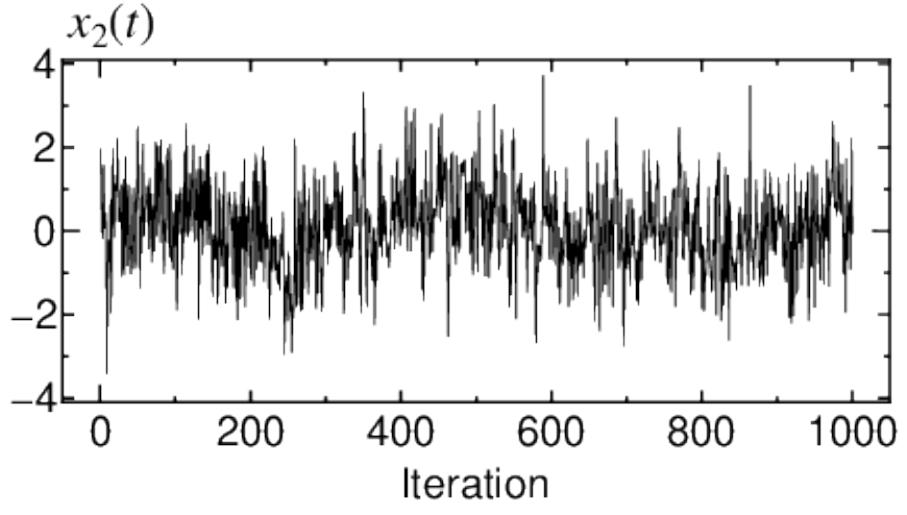}\\
  \includegraphics[width=6.5cm]{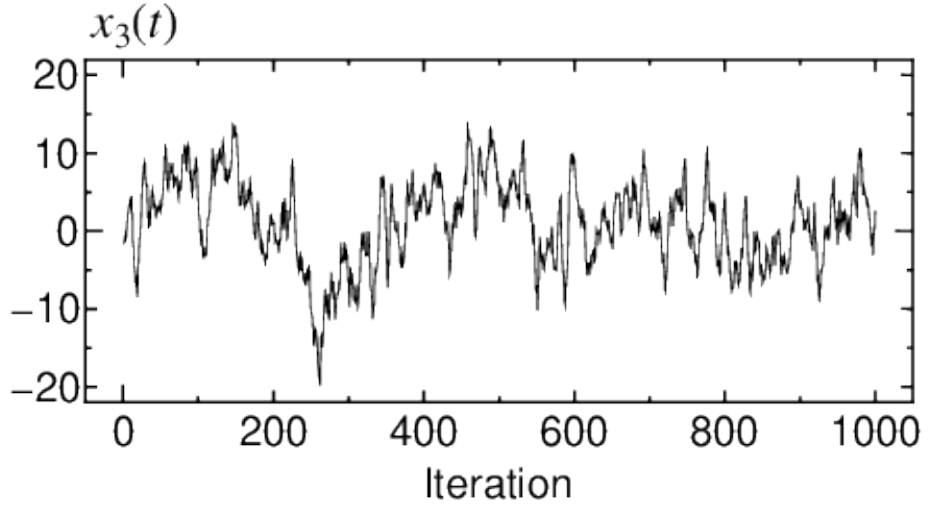}
~~\includegraphics[width=6.5cm]{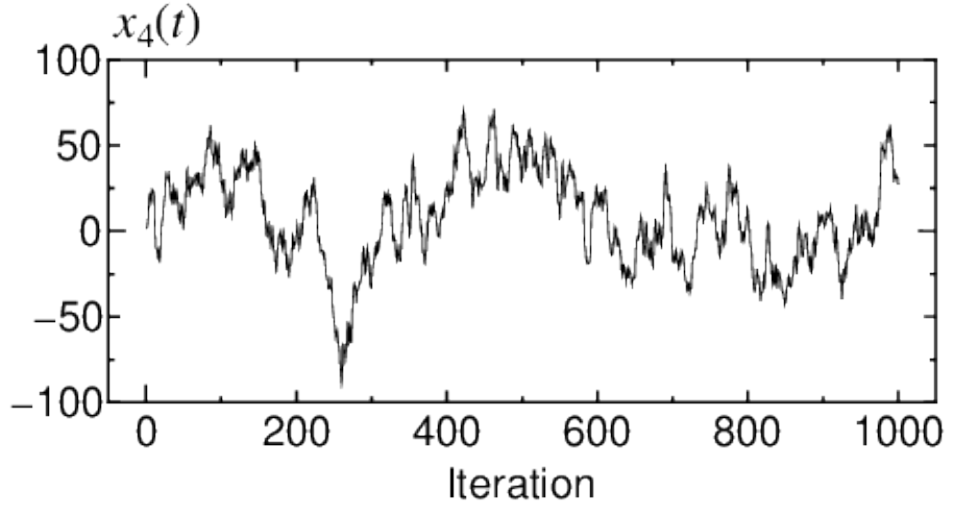}
\caption{Time series data generated by 
         Eqs.~(\ref{eq:multi_linear_model2-1})--(\ref{eq:multi_linear_model2-4}).
         Irregular fluctuations are observed in all variables 
         with different time scales.
         The behaviour of $ x_1 $ is slow, that of $ x_2 $ is fast,
         and that of $ x_3 $ and $ x_4 $ is moderate.}
\label{fig:example2_RAR}
\end{figure}

The behaviours of the four time series generated by these models are 
shown in Fig.~\ref{fig:example1_RAR} and Fig~\ref{fig:example2_RAR}.
The generated data are contaminated by Gaussian observational noise 
with the mean~zero and the standard deviation~0.01.
Figure~\ref{fig:example1_RAR} shows that the behaviours of System~1 show 
irregular fluctuations with similar time scales.
Figure~\ref{fig:example2_RAR} shows that the behaviours of System~2 show 
irregular fluctuations with different time scales.
The behaviour of $ x_1 $ is slow, while $ x_2 $ is fast,
and both $ x_3 $ and $ x_4 $ are moderate.

To construct directed networks from the perspective of 
the underlying linear periodic structures 
we construct the network using summarized information of the models.
The basic idea of the summarized information is as follows.
We distinguish the species of the time series, $ x_i $,
from a value of the species at a specific time, $ x_i(t-l) $,
by the term, ``component'' $ x_i $.
For $ x_i(t-l) $, in contrast, we use the term, ``variable''.
We treat the components as the nodes of the network.

We first consider System~1,
Eqs.~(\ref{eq:multi_linear_model1-1})--(\ref{eq:multi_linear_model1-4}).
Eq.~(\ref{eq:multi_linear_model1-1}) shows that  
the component~$ x_1 $ is influenced by three components, $ x_1 $, $ x_2 $ 
and $ x_4 $.
That is, $ x_1 $ should be connected to $ x_2 $ and $ x_4 $.
Similarly, since Eq.~(\ref{eq:multi_linear_model1-2}) shows that  
$ x_2 $ is driven only by itself, $ x_2 $ has no connection. 
Since Eq.~(\ref{eq:multi_linear_model1-3}) shows that $ x_3 $ is driven by 
$ x_1 $ and $ x_4 $, $ x_3 $ should be connected to $ x_1 $ and $ x_4 $.
Finally, since Eq.~(\ref{eq:multi_linear_model1-4}) shows that $ x_4 $ is driven 
by $ x_1 $ and $ x_4 $, $ x_4 $ should be connected to $ x_1 $.
The whole relationship in terms of connectivity is represented
by the following set of reduced expressions:
\begin{align}
 x_1 &= f_1(x_2, x_4),  \label{eq:connectivity_of_linear_model1-1} \\
 x_2 &= 0,              \label{eq:connectivity_of_linear_model1-2} \\
 x_3 &= f_3(x_1, x_4),  \label{eq:connectivity_of_linear_model1-3} \\
 x_4 &= f_4(x_1),       \label{eq:connectivity_of_linear_model1-4}
\end{align}
where $ f_i $ stands for the function representing connectivity
of the $ i $-th component~$ x_i $
and zero means that there is no connection.
These expressions indicate the essential linear periodic structures
and can be treated as summary information.

The summarized information of System~2 using this idea,
Eqs.~(\ref{eq:multi_linear_model2-1})--(\ref{eq:multi_linear_model2-4}), is 
\begin{align}
 x_1 &= f_1(x_2),  \label{eq:connectivity_of_linear_model2-1} \\
 x_2 &= 0,         \label{eq:connectivity_of_linear_model2-2} \\
 x_3 &= f_3(x_2),  \label{eq:connectivity_of_linear_model2-3} \\
 x_4 &= f_4(x_2).  \label{eq:connectivity_of_linear_model2-4}
\end{align}

\begin{figure}[!t]
\centering
(a)\includegraphics[width=6.0cm]{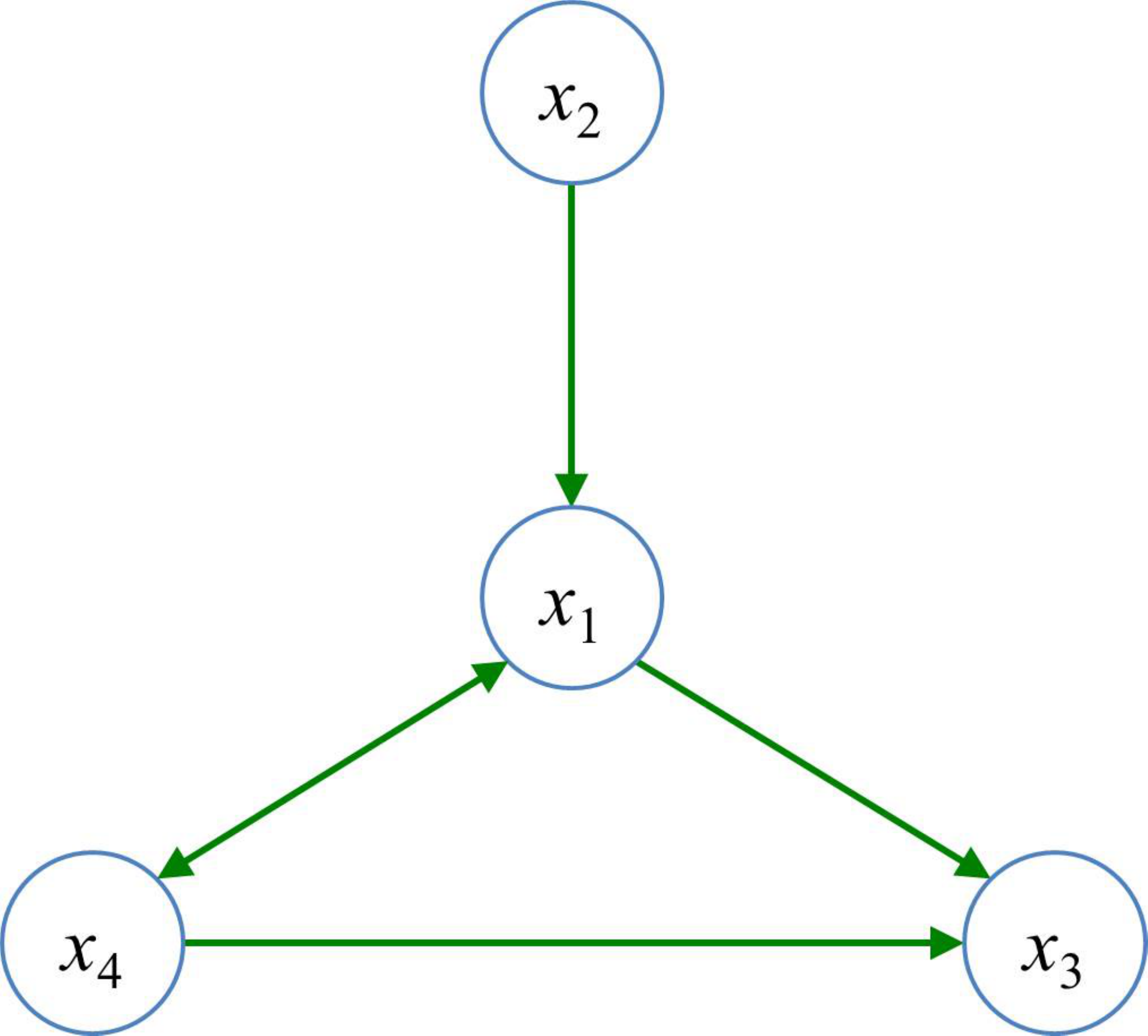}
~~(b)\includegraphics[width=6.0cm]{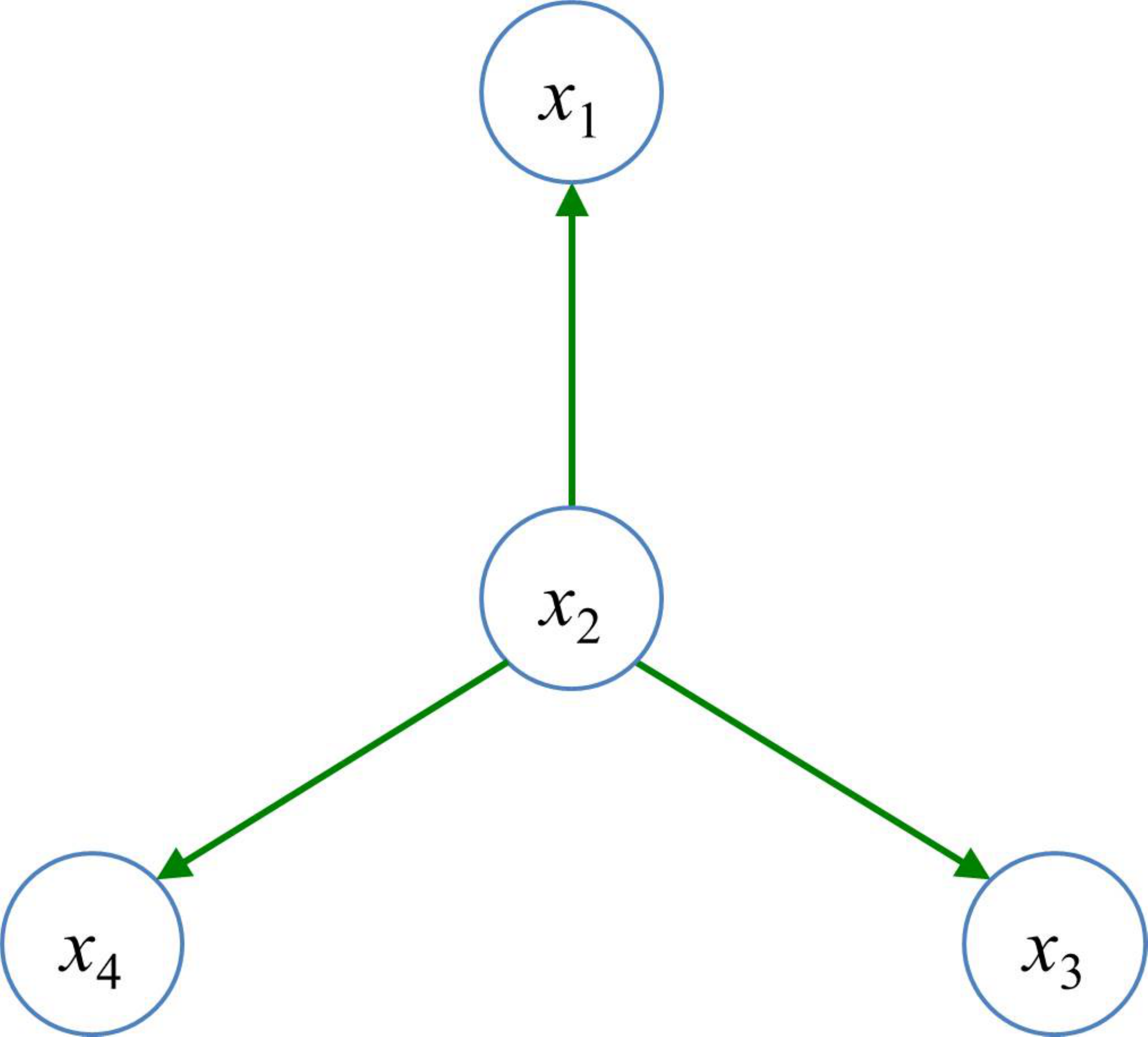}
\caption{(Colour online)~The network and the directed connectivity:
         (a)~the first system, 
         Eqs.~(\ref{eq:connectivity_of_linear_model1-1})--(\ref{eq:connectivity_of_linear_model1-4}),
         and (b)~the second system,
         Eqs.~(\ref{eq:connectivity_of_linear_model2-1})--(\ref{eq:connectivity_of_linear_model2-4}).}
\label{fig:networks_of_known_systems}
\end{figure}

The directed networks constructed based on these summary information 
are shown in Fig.~\ref{fig:networks_of_known_systems}.
As the summarized information reflects the underlying linear periodic structures
of the system, we consider that the networks are not peculiar but reasonable 
and natural.
Hence, when we obtain multivariate time series
we like to construct networks provided with these properties.

There is one point to be mentioned.
One may think that the time delay information is lost in the network, 
as the delays themselves are not encoded in the links.
However, this is not necessarily true in our opinion.
The connections reflect some~(possibly unknown) time delays,
and the time delay connectivity structure is still indirectly encoded 
in the network, but the strength and particular value of time delays are 
not retained indeed.

As shown in Fig.~\ref{fig:example1_RAR},
the behaviours of System~1 generated by 
Eqs.~(\ref{eq:multi_linear_model1-1})--(\ref{eq:multi_linear_model1-4})
show irregular fluctuations and it is difficult to know the relationship 
among the data by visual inspection.
On the other hand, as shown in Fig.~\ref{fig:example2_RAR},
the behaviours of $ x_3 $ and $ x_4 $ in System~2 seem to be 
very similar.
However, this similarity is deceptive.
As Eqs.~(\ref{eq:connectivity_of_linear_model2-3}) and 
(\ref{eq:connectivity_of_linear_model2-4}) indicate, 
although both of $ x_3 $ and $ x_4 $ are influenced by $ x_2 $,
$ x_4 $ is not included in the function of $ x_3 $,
and $ x_3 $ is not included in the function of $ x_4 $.
That is, there is no direct connection to between $ x_3 $ and $ x_4 $.

In the next section we describe fundamental problems with the current approaches
and show that the current approaches fail to construct the networks
shown Fig.~\ref{fig:networks_of_known_systems}.
\section{Fundamental problems with the current approaches}
\label{sec:problems_with_current_approaches}

The directed transfer function~(DTF) and
the partial directed coherence~(PDC) are widely accepted methods 
as the frequency-based approach, and the naive method utilizing 
the cross correlation~(CC) function is extensively used method 
of time-based approach.

In the first place, we discuss about using the multivariate 
auto-regressive (MVAR) model.
Both DTF and PDC are based on the detection of phase differences between 
two signals.
The reliability of the detection obviously depends on the quality of 
the MVAR model.
However, MVAR model cannot meet the expectation.
When building a MVAR model
the conventional strategy is to increase the time delay of all variables 
progressively~\cite{Judd-Mees:1995}.
The optimal model is the one that has the smallest value of a chosen 
information criterion among many models~\cite{Kitagawa-Akaike:modelling78}.
In this strategy, a new term is always added to the previous model at each step,
irrespective of whether the new term is indeed necessary or not.
Furthermore, a set of multivariate time series might simultaneously include 
both short term~(high frequency) and long term~(low frequency) effects.
To treat such multivariate time series appropriately,
the model must include terms with delays of separated time scales
corresponding to each effect.
This is the so-called embedding problem~\cite{Judd-Mees:1998}.
This means that MVAR model often does not contain necessary terms which reflect 
peculiarities of time series.

The fact that
all variables with the same time delays are included in MVAR models,
irrespective of whether some of them are necessary or not,
can cause a problem in parameter estimation, especially
when there is strong collinearity among terms.
In this case,
some parameters have very large or small values~\cite{Nakamura-etal:2003}.
The value of DTF and PDC is obtained using the coefficient of 
MVAR models~\cite{Kaminski-Blinowska:DTF1991,%
Kaminski-etal:DTF2001,Baccala-Sameshima:PDF2001,Astolfi-etal:2007}.
Using unreliable coefficients might cause a serious problem.
One of the main purposes to adopt statistical modeling technique 
is to extract the underlying nature of the data.
However, the MVAR model is obviously inappropriate in this respect.
Since DTF and PDC are based on MVAR modeling, the results
obtained by these methods might be unreliable in some cases.

Next, we discuss about using cross correlation~(CC).
To determine whether two nodes should be connected,
the CC function is used in the naive method.
In this approach it is basically expected that there are some sort of direct 
influence between two signals when these are similar.
Then, it is  considered that ``similarity'' is equivalent to ``relationship''
and ``no similarity'' is equivalent to ``no relationship.''
This idea does not always work well~\cite{Nakamura-etal:SSS_network2016}.
The CC function is only a useful measure of {\em linear} similarity.
More precisely, the CC function is inherently a useful statistic 
to examine a rectilinear proportional relationship between two signals.
However, even if the superposition principle is assumed, 
a rectilinear proportional relationship among some data is not always retained.
As experimental time series will typically show some irregular fluctuations, 
the CC function will often be insufficient. 

Finally, the frequency-based and the time-based approaches both need 
a threshold value to determine the existence of relationship.
It is difficult to select an appropriate threshold
and to determine the precise relationship 
among the various variables by a threshold value.

We explicitly show that all of DTF, PDC and the naive method
fail for two systems with linear periodic structures,
Eqs.~(\ref{eq:multi_linear_model1-1})--(\ref{eq:multi_linear_model1-4})
and Eqs.~(\ref{eq:multi_linear_model2-1})--(\ref{eq:multi_linear_model2-4}),
which we introduced in Section~\ref{sec:desired_networks}.

\subsection{Failure of the current approaches}
\label{sec:cases_the_current_approaches_do_not_work}

Eqs.~(\ref{eq:multi_linear_model1-1})--(\ref{eq:multi_linear_model1-4}) 
of System~1 generate time series with similar time scales
and Eqs.~(\ref{eq:multi_linear_model2-1})--(\ref{eq:multi_linear_model2-4}) 
of System~2 generate time series with different time scales.
We use 1000~data points and the data are contaminated 
by Gaussian observational noise with the mean~zero and 
the standard deviation~0.01.
We apply DTF, PDC and the naive method to the data generated by both
systems\footnote{We generated another four sets of time series and 
    applied the naive method to the data.  We show typical result in each case.}
and examine the existence of relationships according to usual procedures described
in Section~\ref{sec:how_to_determine_relationships}.

To apply DTF and PDC
an appropriate MVAR model is necessary.
Such a MVAR model is usually determined by Akaike Information Criterion~(AIC) or 
Schwarz Information Criterion~(SIC)~\cite{Toppi-etal:2012}.
The SIC formula is defined by
    \begin{eqnarray}
        \mathrm{SIC}(k) = n \ln \frac{ {\bf e}^T {\bf e} }{n} + k \ln n,
    \label{eq:SIC_eq}
    \end{eqnarray}
where $ n $ is the number of data points, $ k $ is the model size,
and $ \bf e $ is the fitting errors\footnote{The SIC is 
    also known as the Bayesian Information Criterion~(BIC) and description length 
    proposed by Rissanen has essentially the same 
    formula~\cite{Nakamura-etal:IC2006}.}~\cite{Schwarz:SIC}.
The best model selected by such an information criterion is treated as 
the appropriate MVAR model.
In this paper we use SIC to find appropriate MVAR models.
\subsubsection{System~1: the case of similar time scales}
\label{sec:case:_similar time scales}

\begin{figure}[!h]
\centering
(a)\includegraphics[width=6.0cm]{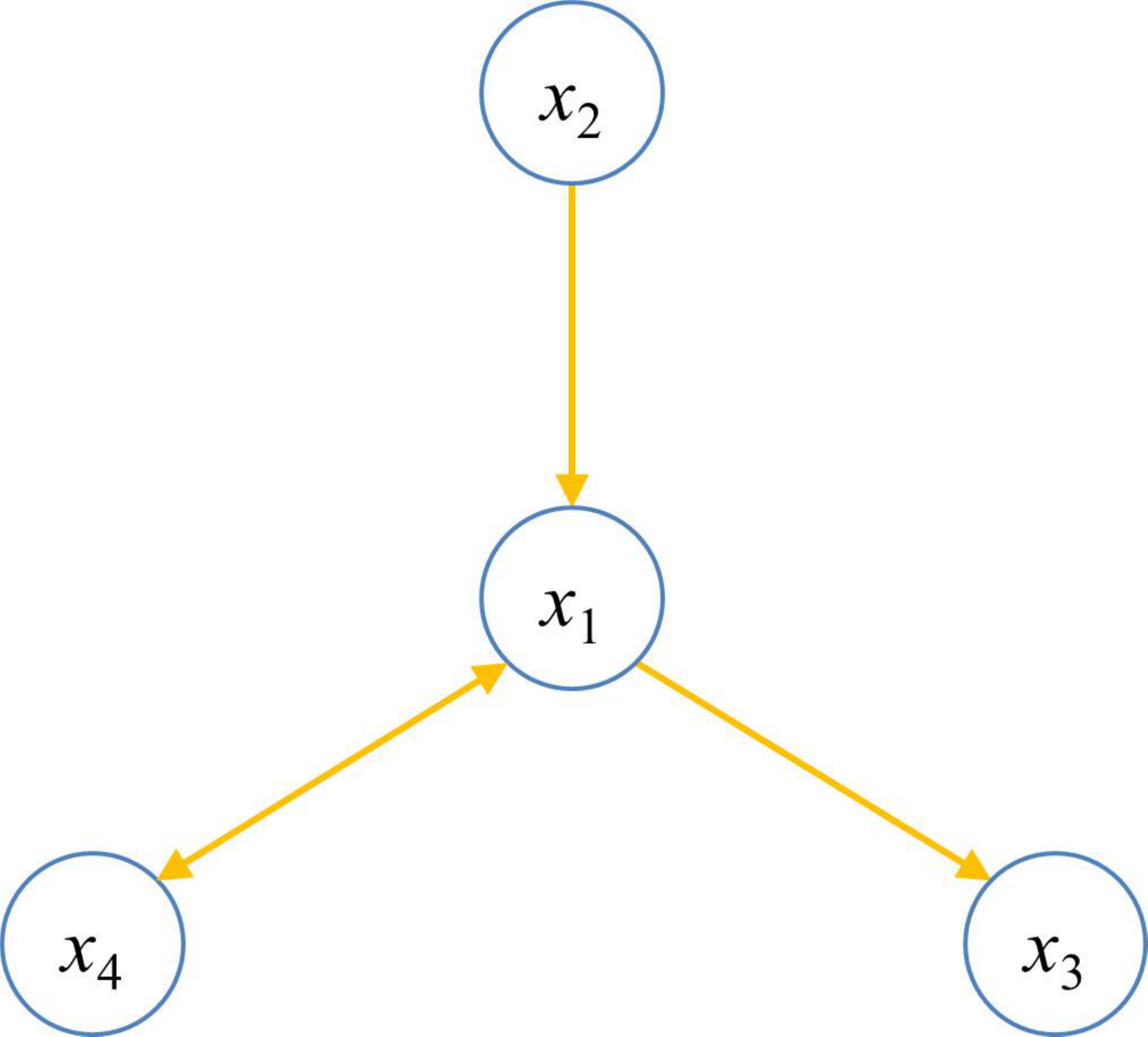}
~~(b)\includegraphics[width=6.0cm]{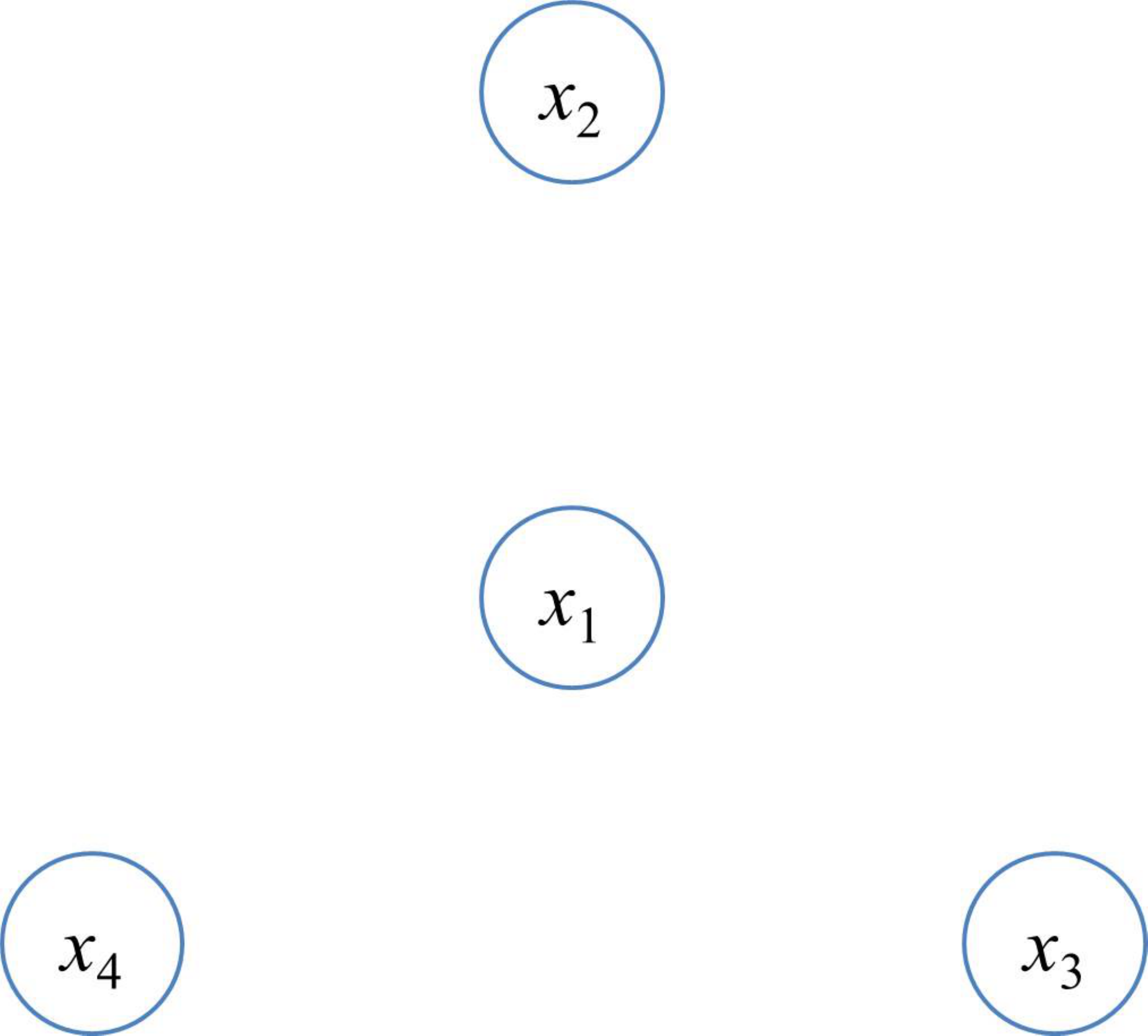}
\caption{(Colour online)~The linkage of constructed network using 1000~data points
         generated by generated by 
         Eqs.~(\ref{eq:multi_linear_model1-1})--(\ref{eq:multi_linear_model1-4}).
         (a)~the network when DTF and PDC are applied to the data,
         and (b)~the network constructed by the naive method using the CC with 
         the threshold~$ 0.5 $,
         The values of the CC are shown in Table~\ref{tab:CC_of_toy_model1}.
         The network constructed by DTF and PDC are the same.}
\label{fig:network_of_toy_model1}
\end{figure}

We first use data generated by 
Eqs.~(\ref{eq:multi_linear_model1-1})--(\ref{eq:multi_linear_model1-4}).
When DTF and PDC are applied to the data,
the size of the best MVAR model is three.
As Eqs.~(\ref{eq:multi_linear_model1-1})--(\ref{eq:multi_linear_model1-4}) show,
the largest time delays for $ x_1 $, $ x_2 $, $ x_3 $ and $ x_4 $ are
$ 3 $, $ 6 $, $ 0 $ and $ 9 $, respectively.
Hence, this indicates that the best MVAR model does not cover the time delays of 
$ x_2 $ and $ x_4 $.
The summarized information obtained by DTF and PDC is
$ x_1 = f_1(x_2, x_4) $, $ x_2 = 0 $, $ x_3 = f_3(x_1) $ and $ x_4 = f_4(x_1) $.
Table~\ref{tab:CC_of_toy_model1} shows all the values of the CC function
and all the values are smaller than the threshold value~$ 0.5 $.
Based on the results we construct the networks.
Figure~\ref{fig:network_of_toy_model1} shows that 
the networks constructed by DTF, PDC and the naive method
are different from network shown in Fig.~\ref{fig:networks_of_known_systems}(a).
Fig.~\ref{fig:network_of_toy_model1}(a) shows the network constructed by
DTF and PDC.
Although there is no link between $ x_3 $ and $ x_4 $,
other directed links are the same as that shown 
in Fig.~\ref{fig:networks_of_known_systems}(a).
However, Fig.~\ref{fig:network_of_toy_model1}(b) shows that
there is no link among the nodes in the network constructed
by the naive method.
%
\begin{table}[!t]
\caption{The largest absolute values of the CC function 
         of all possible pairs between the time lag $ -10 $ and $ 10 $,
         where the number in the parentheses is the time lag 
         when the CC function has the largest absolute value.
         The data are generated by 
         Eqs.~(\ref{eq:multi_linear_model1-1})--(\ref{eq:multi_linear_model1-4}),
         and the values are estimated using 1000 data points.}
\begin{center}
\footnotesize
\begin{tabular}{|c|c|c|c|c|}
\hline
   & ~~~~~~~$ x_1 $~~~~~~~ & ~~~~~~~~$ x_2 $~~~~~~~~ & ~~~~~~~~$ x_3 $~~~~~~~~ & ~~~~~~~~$ x_4 $~~~~~~~~ \\
\hline
~~~~$ x_1 $~~~~        & 1.0000     & ---  & --- & --- \\
\hline
$ x_2 $ & 0.4109 (-4) & 1.0000     & --- & --- \\
\hline
$ x_3 $ & 0.3293 (2)  & 0.0782 (5) & 1.0000 & --- \\
\hline
$ x_4 $ & 0.3221 (2)  & 0.1240 (6) & 0.3919 (-9) & 1.0000 \\
\hline
\end{tabular}
\end{center}
    \label{tab:CC_of_toy_model1}
\end{table}
\subsubsection{System~2: the case of distinct time scales}
\label{sec:case:_different_time_scales}

We next use data generated by 
Eqs.~(\ref{eq:multi_linear_model2-1})--(\ref{eq:multi_linear_model2-4}).
When DTF and PDC are applied to the data,
the size of the best MVAR model is four.
As Eqs.~(\ref{eq:multi_linear_model2-1})--(\ref{eq:multi_linear_model2-4}) show,
the largest time delays for $ x_1 $, $ x_2 $, $ x_3 $ and $ x_4 $ are
$ 4 $, $ 10 $, $ 1 $ and $ 1 $, respectively.
Hence, this indicates that the best MVAR model does not cover the time delay 
of $ x_2 $.
The summarized information obtained by DTF is
$ x_1 = f_1(x_3) $, $ x_2 = f_2(x_4) $, $ x_3 = f_3(x_1, x_2, x_4) $ and 
$ x_4 = f_4(x_2) $, and that obtained by PDC is $ x_1 = f_1(x_2) $, 
$ x_2 = f_2(x_4) $, $ x_3 = f_3(x_2) $ and $ x_4 = f_4(x_2) $.

\begin{figure}[!h]
\centering
(a)\includegraphics[width=6.0cm]{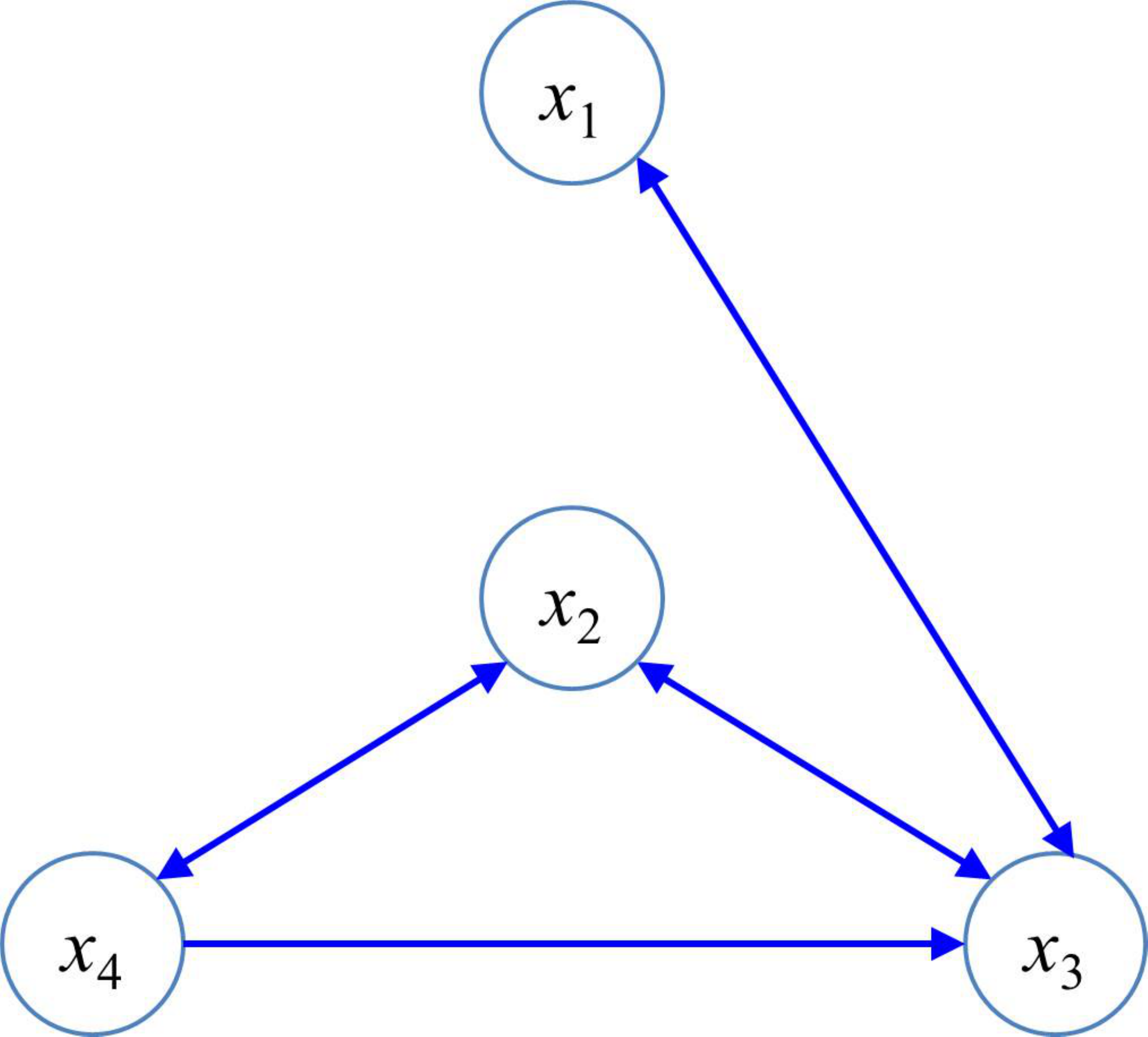}
~~(b)\includegraphics[width=6.0cm]{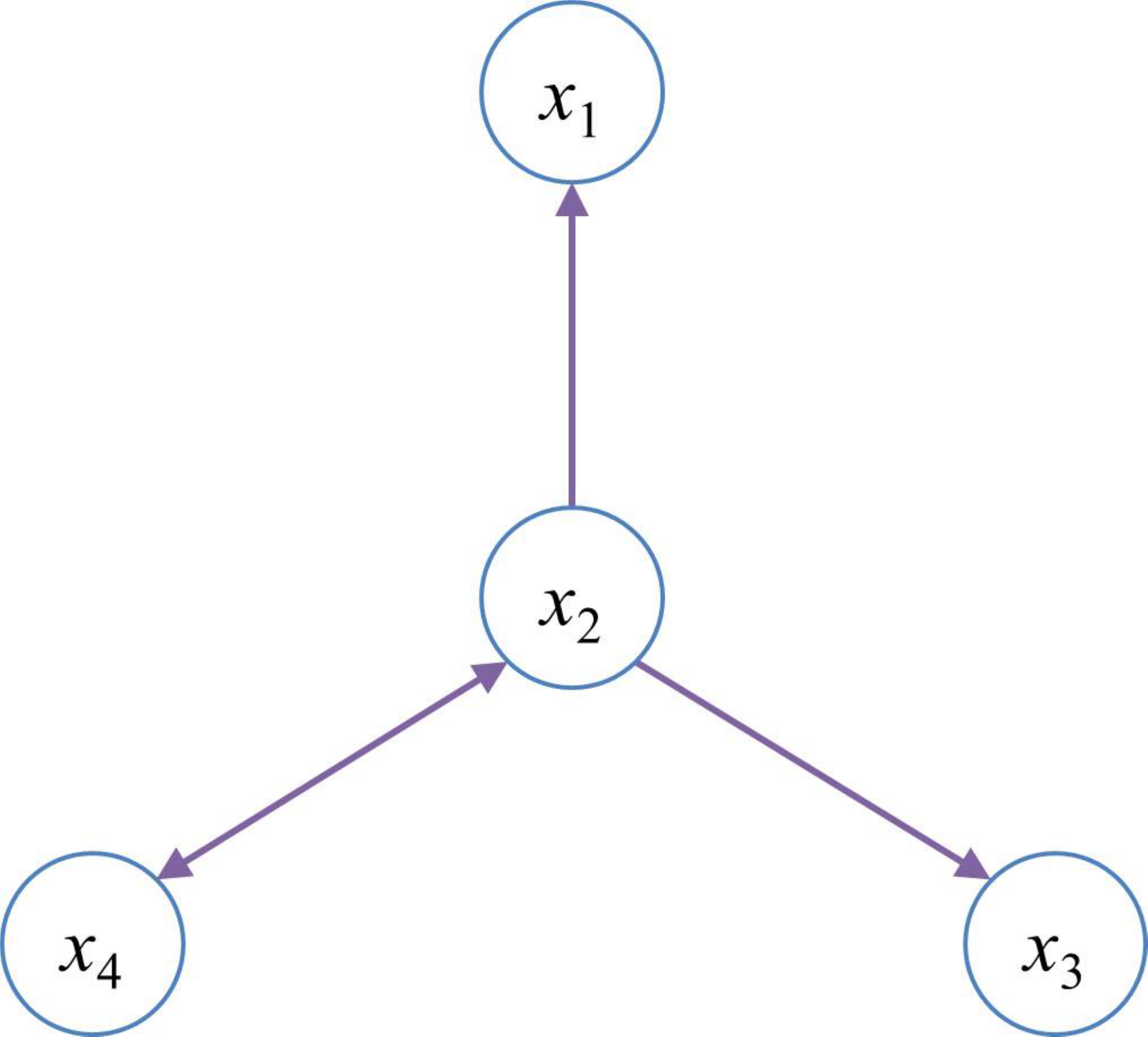}\\
\vspace{0.5cm}
(c)\includegraphics[width=6.0cm]{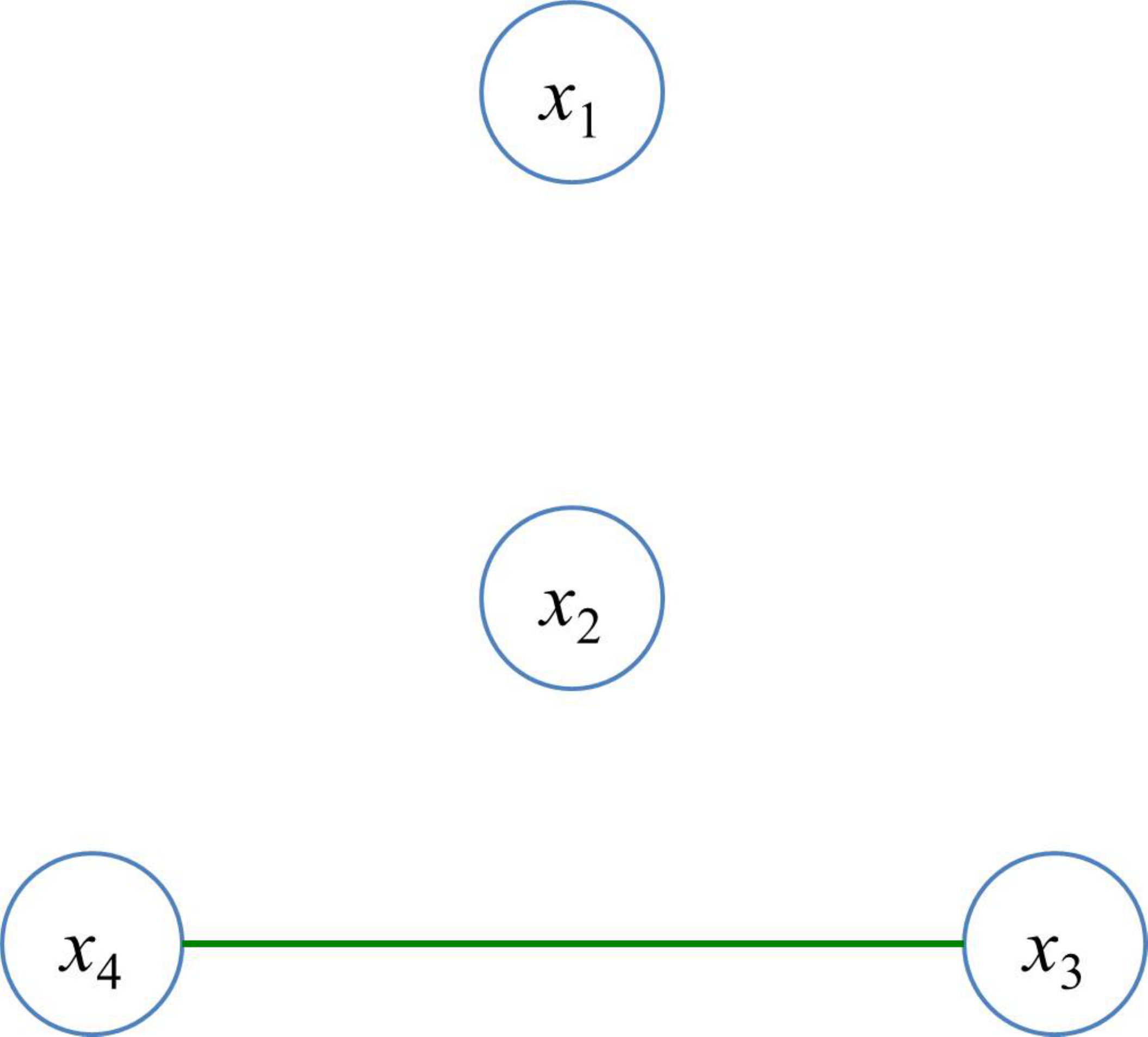}
\caption{(Colour online)~The linkage of constructed network using 1000~data points
         generated by generated by 
         Eqs.~(\ref{eq:multi_linear_model2-1})--(\ref{eq:multi_linear_model2-4}).
         (a)~the network when DTF is applied to the data,
         (b)~the network when PDC is applied to the data,
         and (b)~the network constructed by the naive method using the CC 
         with the threshold~$ 0.5 $,
         The values of the CC are shown in Table~\ref{tab:CC_of_toy_model2}.}
\label{fig:network_of_toy_model2}
\end{figure}

Table~\ref{tab:CC_of_toy_model2} shows all the values of the CC function.
Based on the results we construct the networks.
Figure~\ref{fig:network_of_toy_model2} shows that 
the networks constructed by DTF, PDC and the naive method are different from 
the network shown in Fig.~\ref{fig:networks_of_known_systems}(b).
Figure~\ref{fig:network_of_toy_model2}(a) shows that DTF fails to detect
the relationship between $ x_1 $ and $ x_2 $ and creates non-existent links
between $ x_1 $ and $ x_3 $ and between $ x_3 $ and $ x_4 $.
Figure~\ref{fig:network_of_toy_model2}(b) shows that connectivity in the network 
constructed by PDC is the same as that in 
Fig.~\ref{fig:networks_of_known_systems}(b).
However, there is one non-existent directed link from $ x_4 $ and $ x_2 $.
Figure~\ref{fig:network_of_toy_model2}(c) shows that the naive method
fails to detect most of links and creates one non-existent link between $ x_3 $ 
and $ x_4 $.
The value between $ x_3 $ and $ x_4 $ is larger than the commonly used threshold 
value~$ 0.5 $~\cite{Tsonis-Swanson:2008,Tse-etal:2010}.
The behaviour of $ x_3 $ is visually similar to that of $ x_4 $ 
as shown in Fig.~\ref{fig:example2_RAR}.
Hence, it seems likely that there is direct relationship between them.
However, it is clearly untrue.
%
\begin{table}[!t]
\caption{The largest absolute values of the CC function 
         of all possible pairs between the time lag $ -10 $ and $ 10 $,
         where the number in the parentheses is the time lag 
         when the CC function has the largest absolute value.
         The data are generated by 
         Eqs.~(\ref{eq:multi_linear_model2-1})--(\ref{eq:multi_linear_model2-4}),
         and the values are estimated using 1000 data points.}
\begin{center}
\footnotesize
\begin{tabular}{|c|c|c|c|c|}
\hline
   & ~~~~~~~$ x_1 $~~~~~~~ & ~~~~~~~~$ x_2 $~~~~~~~~ & ~~~~~~~~$ x_3 $~~~~~~~~ & ~~~~~~~~$ x_4 $~~~~~~~~ \\
\hline
~~~~$ x_1 $~~~~        & 1.0000     & ---  & --- & --- \\
\hline
$ x_2 $ & 0.0903 (-10) & 1.0000     & --- & --- \\
\hline
$ x_3 $ & 0.1497 (10)  & 0.4693 (5) & 1.0000 & --- \\
\hline
$ x_4 $ & 0.2322 (-10) & 0.4969 (4) & 0.7729 (-1) & 1.0000 \\
\hline
\end{tabular}
\end{center}
    \label{tab:CC_of_toy_model2}
\end{table}
\subsubsection{Results of the current approaches}
\label{sec:some_observations}

As mentioned in Section~\ref{sec:problems_with_current_approaches}
MVAR model cannot always include terms with delays of separated time scales.
In Sections~\ref{sec:case:_similar time scales} and 
\ref{sec:case:_different_time_scales} we found that the best MVAR models 
for both cases do not cover the largest time delay of some components.
We are afraid of that this problem may happen at any moment.
Even if the MVAR models do not contain necessary terms,
it is acceptable if the constructed networks based on the MVAR models
are the same as desired networks.
However, the current approaches unfortunately cannot construct desired 
networks for the two examples.\footnote{We use 
    {\it p}-value to determine the existence of relationships 
    when applying DTF and PDC.
    Although we apply the false discovery rate~(FDR) correction 
    and Monte Carlo hypothesis testing to the results~\cite{Toppi-etal:2012,%
    Theiler-Prichard:Monte_Carlo_test}, the networks are different from that 
    shown Fig.~\ref{fig:networks_of_known_systems}.}

We consider that there are at least four types of relationships
for the elements of the assumed system:
(i)~unilateral influence from an element~$ A $ to an element~$ B $,
(ii)~unilateral influence from $ B $ to $ A $,
(iii)~mutual influence between $ A $ and $ B $,
and (iv)~influence to $ A $ and $ B $ from an independent~(third-party) 
element~$ C $.
It is preferable to be able to distinguish these relationships.
In this respect, Zalesky~{\it et al.} have pointed out that the CC function 
should be used cautiously in network construction~\cite{Zalesky-etal:2012}.
In our opinion, this point has been overlooked by the user of the naive method.

Hence, the following four points are particularly important 
in constructing networks for multivariate time series:
(1)~multivariate time series may include both short time and long time effects,
whose time scales are well separated,
(2)~we should be able to identify linear periodic structures included 
in multivariate data,
(3)~we should not be deceived by apparent behaviours of data,
and (4)~relationships among data should not be determined by
an externally selected value of the threshold.
Although these points are mentioned separately, 
these are strongly interconnected when constructing networks from multivariate 
time series.
Furthermore, it will be preferable to construct directed networks.
To fulfil these requirements and overcome the drawbacks with 
the current approaches
we propose a method based on an information theoretic reduction of 
a linear~(auto-regressive) model for multivariate time series.

\section{An approach based on the reduced auto-regressive model}
\label{sec:new_approach}

As indicated in Section~\ref{sec:desired_networks},
if we have enough information about the exact dynamical equations of the system,
the faithful network representation can be obtained from the summarized information.
Unfortunately, it is usually difficult to obtain the information in practice.
In most cases, we have to start only from observed data without the knowledge of 
the underlying dynamical system.
Hence, the main issue is to obtain the dynamical relationship among components 
as faithful as possible only from the observed multivariate time series data.
We first consider this problem in cases where a perfect linear model of 
a system exists.
We use Systems~1 and 2 introduced in Section~\ref{sec:desired_networks},
where the current approaches do not work well.
In Section~\ref{sec:imperfect_model_scenario} we will consider this problem
in uncertain situations, when no correct linear system of a system exists.
\subsection{Reduced auto-regressive~(RAR) model}
\label{sec:RAR_model}

To precisely identify the underlying linear periodic structures for multivariate 
time series we apply an information theoretic reduction of linear models, 
the reduced auto-regressive~(RAR) model~\cite{Judd-Mees:1995,Judd-Mees:1998}.
There are strong information theoretic arguments to support that 
RAR model can detect any periodicities built into a given time 
series~\cite{Small-Judd:1999}.
The RAR model includes terms only when their combination
  contributes significantly to the model as an entire system
  in terms of a suitably chosen information criterion~\cite{Judd-Mees:1998}
  and allows to contain terms with short and large time delays concurrently
  unlike the AR model,
  even when the time scales of the delays are completely
  different.\footnote{When unit time delay in AR model is necessary, 
    terms with unit time delay are included in RAR model.
    Also, when building RAR model the meaning or role or weight in the RAR model
    is not checked.}
The RAR model is basically composed of the terms with time 
intervals~(periodicities) at which underlying or characteristic 
situations~(behaviours) are sharply or clearly repeated in the data, 
irrespective of whether the patterns or data are linear or 
nonlinear \footnote{The cross correlation function can examine 
    a rectilinear proportional relationship between two signals,
    irrespective of whether the two signals are linear or nonlinear.}.
The RAR model is thus effective in modelling both linear and 
nonlinear data~\cite{Judd-Mees:1995,Judd-Mees:1998,Small-Judd:1999}.
We explicitly show in \ref{sec:detection_by_RAR_model} that 
the RAR model can identify periodicities in both linear and nonlinear data.

An RAR model from given a univariate~(scalar) time series is constructed 
as follows.
Given a univariate time series $ \{ x(t) \}_{t=1}^n $ of $ n $~observations,
an RAR model with the largest time delay~$ l_w $ is expressed as
\begin{align}
    x(t) &= a_{0} + a_{1} x(t - l_{1}) + a_{2} x(t - l_{2}) + \dots + a_{w} x(t - l_{w}) + \varepsilon(t) \notag \\
         &= a_0 + \sum_{i=1}^{w} a_i x(t-l_i) + \varepsilon(t),
    \label{eq:RAR}
\end{align}
where $ 1 \leq l_1 < l_2 \cdots < l_w $, 
$ a_i \; (i = 0, 1, 2 \dots, w) $ are parameters to be determined, 
and $ \varepsilon(t) $ is assumed to be independent and identically
distributed Gaussian random variables, which are interpreted as fitting errors.
The parameters~$ a_i $ are chosen to minimize the sum of the squares of 
fitting errors.
To build an RAR model we prepare candidate basis functions used in the modelling, 
in the form of a dictionary, and select the most appropriate basis functions that 
can extract the temporal structure of the time series.
The number of candidate basis functions included in a dictionary
is not restricted {\it a priori}.
When we assume a linear model
the basis functions are a constant and linear terms.
For selecting basis functions, various algorithms have been proposed,
which are proven to be effective in modelling both linear and nonlinear dynamics.
The models obtained by these algorithms are considered to be
nearly optimal~\cite{Judd-Mees:1995,Judd-Mees:1998,Small-Judd:1999,
Nakamura-etal:2004,Nakamura-Small:2006}.
In this paper, we adopt a selection algorithm using the total 
error~\cite{Nakamura-etal:2004},
which will be described later in this subsection.

It is straightforward to apply this methodology to multivariate time series.
A set of multivariate RAR models is expressed by
\begin{equation}
    x_i(t) = a_{i,0} + \sum_{j=1}^{N} \left( \sum_{k=1}^{w_j} a_{i,j,k} \; x_j(t - l_k) \right)
             + \varepsilon_i(t) \;\;\; (i = 1, 2, \dots, N),
    \label{eq:multivariate_RAR}
\end{equation}
where $ N $ is the number of components and
$ l_{w_i} (\geq 0) $ is the largest time delay of the $ i $-th component.

\subsection{How to find an optimal RAR model}
\label{sec:find_optimal_RAR_model}

In what follows, an information criterion approach is used to evaluate 
and obtain the best~(optimal) model among many. 
The model that gives the minimum of the information criterion 
is considered to be the best model~\cite{Akaike:AIC74,Kevin:book_chapter2003}.
Various information criteria have been proposed for 
their own purposes~\cite{Judd-Mees:1995,Schwarz:SIC,Akaike:AIC74,Rissanen_book,%
Rissanen:NML}. 
For determining the best model we adopt the Description Length~(DL) suitably 
modified by Judd and Mees~\cite{Judd-Mees:1995},
because the DL modified by Judd and Mees has proven to be effective in modelling 
nonlinear dynamics~\cite{Judd-Mees:1998,Small-Judd:1999}, 
and it has fewer approximations than other information criteria, 
though slightly more calculations are needed~\cite{Judd-Mees:1995}.
Hence, the DL is more reliable for the present purpose~\cite{Nakamura-etal:IC2006}.

When the $ \varepsilon(t) $ and $ \varepsilon_i(t) $ 
in Eqs.~(\ref{eq:RAR}) and~(\ref{eq:multivariate_RAR}) 
are assumed to be Gaussian and the $ a_i $ and $ a_{i,j,k} $ 
in Eqs.~(\ref{eq:RAR}) and~(\ref{eq:multivariate_RAR}) have been chosen 
to minimise the sum of squares of the prediction errors~$ {\bf e} = y - \hat{y} $ 
where $ y $ is the observational data and $ \hat{y} $ is the predicted data,
Judd and Mees show that the description length is bounded by

\begin{eqnarray}
    DL(k) = \left( \frac{n}{2} - 1 \right) \ln
    \frac{{\bf e}^T {\bf e}}{n} + \left( k + 1 \right)
    \left( \frac{1}{2} + \ln \gamma \right) - \sum_{i=1}^{k} \ln \delta_i,
    \label{eq:DL}
\end{eqnarray}
where $ n $ is the length of the time series to be fitted,
$ k $ is the number of parameters~(or model size), 
$ \gamma $ is related to the scale of the data, 
and the variables $ \delta $ can be interpreted as the relative precision
to which the parameters are specified.
The factor $ \gamma $ is a constant and typically fixed to be 
$ \gamma = 32 $~\cite{Judd-Mees:1995}.
The first term in the description length equation, 
$ \left( \frac{n}{2} - 1 \right) \ln \frac{{\bf e}^T {\bf e}}{n} $, 
is the penalty for the model prediction errors 
and is derived from the conventional log-likelihood expression. 
In the case of DL that derivation is a little circuitous 
as the DL penalty is measuring the cost of encoding those errors.
More thorough arguments for the details of the RAR model and the DL can be found 
in~\cite{Judd-Mees:1995,Judd-Mees:1998}.

To build an RAR model we need to select the optimal subset from a dictionary
of basis functions.
In this paper, we use a selection algorithm using the total error,
because this algorithm is able to obtain better models in most cases 
than others with reasonable computation time~\cite{Nakamura-etal:2004}.
As the bottom-up approach has proven to be effective~\cite{Judd-Mees:1995,%
Judd-Mees:1998,Small-Judd:1999},
we first apply the bottom-up method using the total error~\cite{Nakamura-etal:2004}.
In searching for the best model,
we might be trapped in one of the local minima of the Description Length.
To avoid this situation and to reduce the problem to a manageable size,
a model is also built from the complete dictionary using the top-down method, 
but starting from the model whose size is 10 larger than that of the best model
obtained by the bottom-up method.
This method do no worse than the bottom-up method.
For more details on this procedure and the relevant approaches 
see \cite{Judd-Mees:1995,Judd-Mees:1998,Small-Judd:1999,Nakamura-etal:2004,%
Small-Judd:1998}.
We select the model as the best model whose description length is the smallest
with this procedure.
However, it should be noted that selecting the optimal subset from a dictionary 
is an NP-hard problem that usually has to be solved 
heuristically~\cite{Judd-Mees:1995}.
We will discuss more on the problems or difficulties with building RAR models 
in Section~\ref{sec:reexamination_of_the_proposed_method}.

\subsection{Confirmation of reproducibility}
\label{sec:confirmation_of_reproducibility}

We apply the RAR modelling technique to the data represented 
in Figs.~\ref{fig:example1_RAR} and \ref{fig:example2_RAR} 
to investigate whether we can reconstruct System~1,
Eqs.~(\ref{eq:multi_linear_model1-1})--(\ref{eq:multi_linear_model1-4}),
and System~2,
Eqs.~(\ref{eq:multi_linear_model2-1})--(\ref{eq:multi_linear_model2-4}),
only from the generated data sets.
In this case, we have four time series~(that is, $ x_1(t) $, $ x_2(t) $, 
$ x_3(t) $ and $ x_4(t) $) of 1000 data points with Gaussian observational noise
for each data set.
For Gaussian observational noise,
we use four different noise level with
the standard deviation~0.01, 0.02, 0.05 and 0.1.
All of the mean values are fixed to zero.
For each observational noise level,
we prepare five sets of time series with different noise 
realizations.\footnote{Problems associated 
    with the observational noise are of great significance.
    On the other hand, it is obvious that less observational noise 
    is better for not only the proposed method but generically.
    Moreover, the robustness of any method strongly depend on nature of 
    target time series and systems.
    The proposed method works well when the observational noise level is not
    significant relative to the target system.}
Choosing time delays up to 10 for time series of each component
and the constant function give 41 candidate basis functions 
in the dictionary.\footnote{These are the constant function, 
    $ x_1(t-1) $, $ x_1(t-2) $, $ \dots $, $ x_1(t-10) $,
    $ x_2(t-1) $, $ \dots $, $ x_2(t-10) $,
    $ x_3(t-1) $, $ \dots $, $ x_3(t-10) $,
    and $ x_4(t-1) $, $ \dots $, $ x_4(t-10) $.}
Using the dictionary we build the multivariate RAR model 
for four components, $ x_1 $, $ x_2 $, $ x_3 $, and $ x_4 $.
We show some of the results when the observational noise level is 0.01.
The best models for System~1,
Eqs.~(\ref{eq:multi_linear_model1-1})--(\ref{eq:multi_linear_model1-4}), are 
\begin{align}
 x_1(t) &= 1.1547 + 0.4123 \; x_1(t-1) - 0.2185 \; x_1(t-3) \notag \\
        &\phantom{=} + 0.3148 \; x_2(t-4) + 0.2359 \; x_4(t-7), \label{eq:obtained_RAR1-1} \\
 x_2(t) &= 1.8172 + 0.6541 \; x_2(t-1) - 0.1952 \; x_2(t-6),  \label{eq:obtained_RAR1-2} \\
 x_3(t) &= 2.1602 + 0.2030 \; x_1(t-2) + 0.3102 \; x_4(t-9), \label{eq:obtained_RAR1-3} \\
 x_4(t) &= 1.2227 + 0.2390 \; x_1(t-2) + 0.4879 \; x_4(t-1) \notag \\
        &\phantom{=} - 0.3013 \; x_4(t-3), \label{eq:obtained_RAR1-4}
\end{align}
and the best models for System~2,
Eqs.~(\ref{eq:multi_linear_model2-1})--(\ref{eq:multi_linear_model2-4}),
are 
\begin{align}
 x_1(t) &= 12.1800 + 1.2967 \; x_1(t-1) - 0.3054 \; x_1(t-4) + 0.2217 \; x_2(t-3), \label{eq:obtained_RAR2-1} \\
 x_2(t) &= 0.3227 \; x_2(t-1) + 0.1969 \; x_2(t-6), \label{eq:obtained_RAR2-2} \\
 x_3(t) &= 1.1935 \; x_2(t-4) - 0.6524 \; x_2(t-10) + 0.9441 \; x_3(t-1), \label{eq:obtained_RAR2-3} \\
 x_4(t) &= 5.6114 \; x_2(t-3) + 0.9193 \; x_4(t-1). \label{eq:obtained_RAR2-4}
\end{align}
Eqs.~(\ref{eq:obtained_RAR1-1})--(\ref{eq:obtained_RAR1-4})
and Eqs.~(\ref{eq:obtained_RAR2-1})--(\ref{eq:obtained_RAR2-4})
show that all terms included in 
Eqs.~(\ref{eq:multi_linear_model1-1})--(\ref{eq:multi_linear_model1-4})
and Eqs.~(\ref{eq:multi_linear_model2-1})--(\ref{eq:multi_linear_model2-4})
and only these terms are selected,
which means that the same networks as those shown 
in Fig.~\ref{fig:networks_of_known_systems} is constructed.
When the RAR modelling technique is applied to the data in all of the other cases,
the situations are the same:
only all the terms included in 
Eqs.~(\ref{eq:multi_linear_model1-1})--(\ref{eq:multi_linear_model1-4})
and Eqs.~(\ref{eq:multi_linear_model2-1})--(\ref{eq:multi_linear_model2-4})
are selected.
This reproducibility manifests the strong power of the proposed method 
for identifying necessary terms.

\section{Imperfect model scenario: when no correct linear system of a system exists}
\label{sec:imperfect_model_scenario}

\begin{figure}[!t]
\begin{minipage}{0.4\columnwidth}
\centering
    (a)\includegraphics[width=5.0cm]{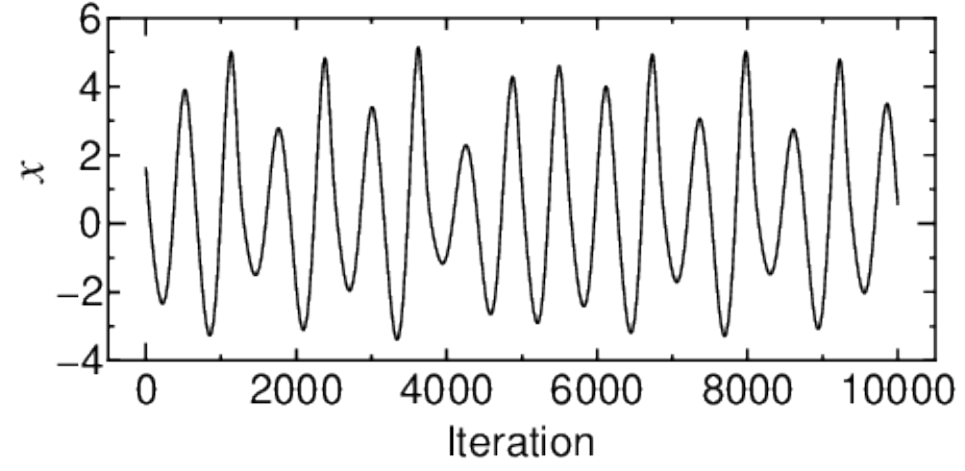}\\
    \vspace{0.1cm}
    (b)\includegraphics[width=5.0cm]{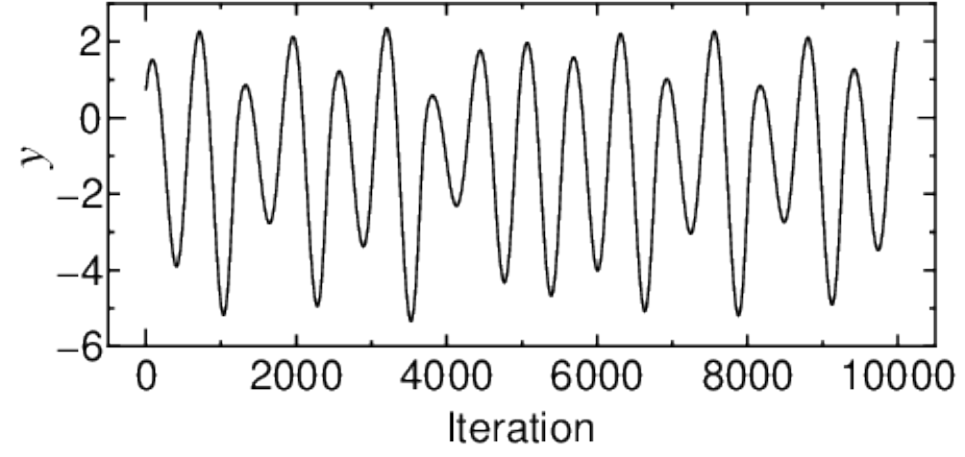}
    \vspace{0.1cm}
    (c)\includegraphics[width=5.0cm]{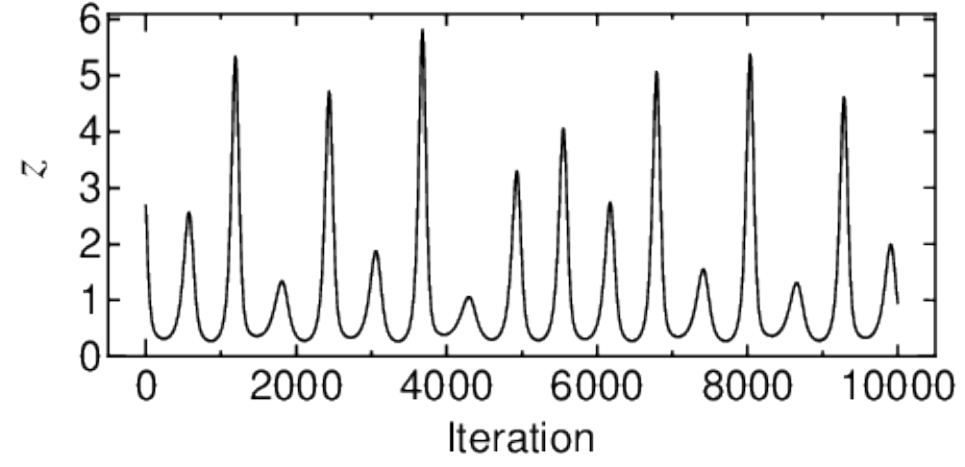}
\end{minipage}
    \begin{minipage}{0.6\columnwidth} 
    \centering
~~~(d)\includegraphics[width=8.0cm]{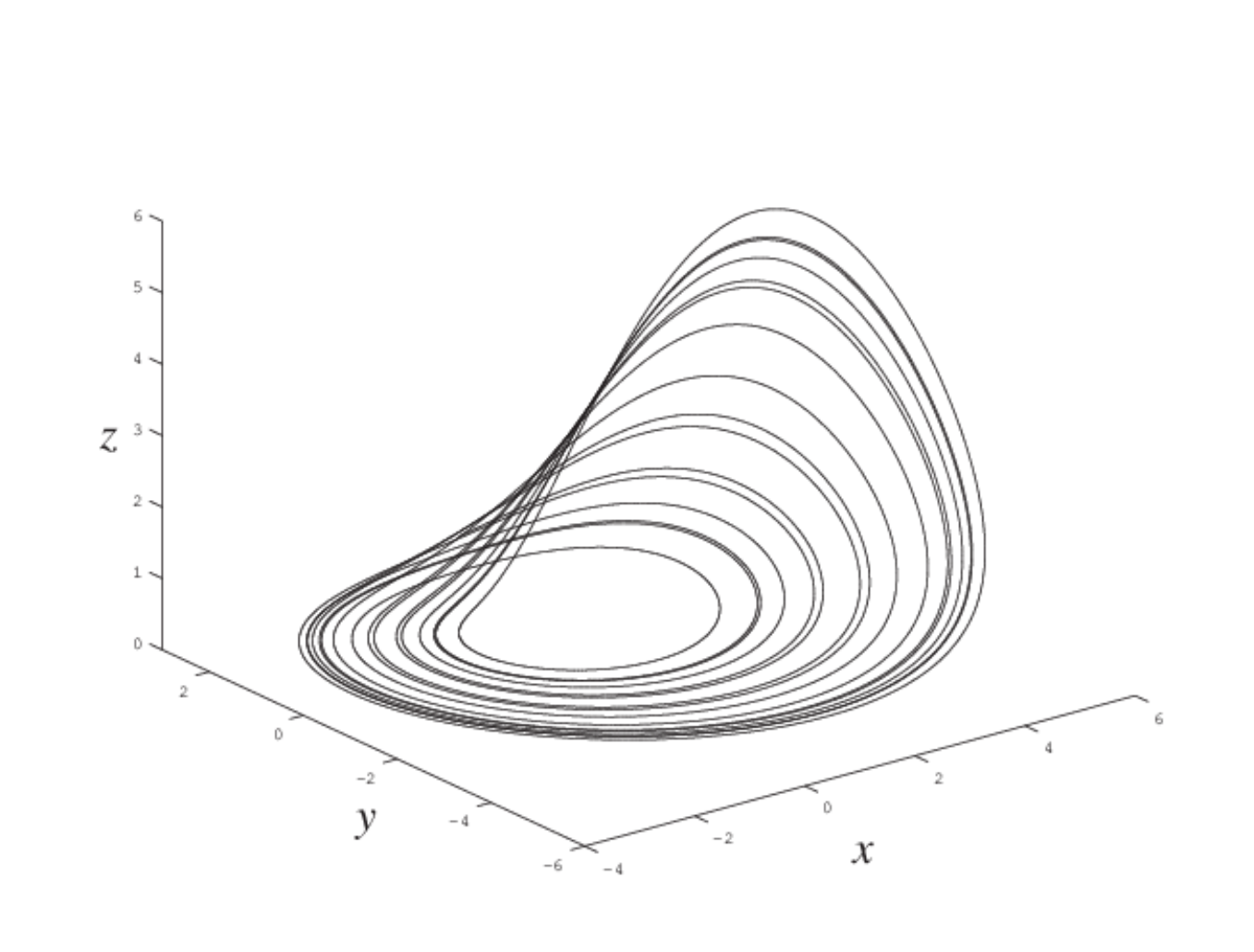} 
\end{minipage} 
\caption{Time series and attractor generated by the R{\"o}ssler systems,
         Eqs.~(\ref{eq:x-roessler_eqs})--(\ref{eq:z-roessler_eqs}),
         where the number of data points is 10~000.
         (a)~$ x $ component, (b)~$ y $ component, (c)~$ z $ component,
         and (d)~attractor composed of $ x $, $ y $ and $ z $.
         Although these time series look periodical behavious as shown 
         in Figs.~\ref{fig:roessler_data}(a)--(c),
         Fig.~\ref{fig:roessler_data}(d) shows that these are chaotic behaviours in reality.}
    \label{fig:roessler_data}
\end{figure}

Thus far we have restricted attention to the perfect model scenario.
The cases we considered were that
correct linear models exist and time series are generated by 
the linear systems including the time delay terms corresponding 
to the periodicities.
That is, there are explicit linear periodic structures among multivariate 
time series.
We confirmed that the RAR model procedure precisely identifies 
the linear periodic structures and then the correct networks are constructed.
However, there may not always be the case,
because even if time series exhibit periodicities~(either exactly periodic and 
nearly periodic behaviour), the system may not contain terms corresponding 
to the periodicities.
In such a case correct linear systems exist no more
or steadfast linear models might not be possible to assume.
Hence, it is important to not build the correct models 
but construct the correct network in this case.
To investigate how the proposed method works in uncertain situations, 
we use the R{\"o}ssler systems presented in the form of a differential equation
as an example.
The equations are given by
\begin{eqnarray}
        dx/dt &=& -y - z,       \label{eq:x-roessler_eqs} \\
        dy/dt &=& x + a y,        \label{eq:y-roessler_eqs} \\
        dz/dt &=& b + z (x - c),  \label{eq:z-roessler_eqs}
\end{eqnarray}
where $ a = 0.398 $, $ b = 2.0 $, $ c = 4.0 $~\cite{Rossler_eqs:1976}.
There is a nonlinear term~$ z \; x $ in Eq.~(\ref{eq:z-roessler_eqs})
and the equations can exhibit chaotic behaviours
when using these parameters~\cite{Small-etal:PPS2001}.
We calculate the equations using the fourth order Runge-Kutta method
with sampling interval~0.01.
As Figs.~\ref{fig:roessler_data}(a)--(c) show,
although time series of each variable is oscillating,
Fig.~\ref{fig:roessler_data}(d) shows that the attractor is chaotic.

Broadly speaking, the meaning of a differential equation is that
a change (or difference) of a variable in a minute time
is expressed by a certain function.
For example, Eq.~(\ref{eq:x-roessler_eqs}) indicates that 
the next value~(or state) of~$ x $ is calculated as the summation of 
the current value of~$ x $ and the minute current value of~$ -y-z $.
That is, although the right side of Eq.~(\ref{eq:x-roessler_eqs}) does not 
contain $ x $, $ x $ is a function composed of $ x $, $ y $ and $ z $ 
in a practical sense.
Hence, $ y $ is a function of $ x $ and $ y $,
and $ z $ is a function of $ x $ and $ z $ in a similar way.

We first investigate the influence of the number of data points.
We use four different data points, 1000, 2000, 5000 and 10~000
and prepare five sets of time series.
These data are contaminated by Gaussian observational noise 
with the mean~zero and the standard deviation~0.01.
As there are three time series, choosing a time delay up to 20 for time series 
of each data and the constant function give 61~candidate basis functions 
in the dictionary.
Using the dictionary we build the multivariate RAR model for each data, 
$ x $, $ y $ and~$ z $.

The summarized information of the obtained three multivariate RAR models 
are shown in Table~\ref{tab:roessler_data_points}.
In all cases the information for $ x $ and $ y $ are correct.
However, the information for $ z $ is different.
When the number of data points is 1000,
the correct information for $ z $ cannot be obtained at all.
When the number of data points is 2000,
although the correct information for $ z $ is obtained,
the wrong information is also obtained.
When the numbers of data points is 5000 and 10~000,
the correct information for $ z $ is always obtained.
We consider that as there is a nonlinear term~$ z \; x $ 
as shown in Eq.~(\ref{eq:z-roessler_eqs}),
the more data points are necessary for $ z $ to extract 
the correct relationship among $ x $, $ y $ and $ z $.
%
\begin{table}[!t]

\caption{Investigation of the influence of the number of data points.
         The components contained in the multivariate RAR models
         using data generated by the R{\"o}ssler systems shown 
         in Fig.~\ref{fig:roessler_data}.
         The numbers of data points are 1000, 2000, 5000 and 10~000,
         five sets of time series are used,
         and the data are contaminated by Gaussian observational noise 
         with the mean~zero and the standard deviation~0.01.
         The value in parentheses is the number of times the result occurred.
         As we use five sets for each data point,
         if the same result is always obtained,
         the value in parentheses is $ 5 $.}
\begin{center}
\footnotesize
\begin{tabular}{| c | c | c | c | c | c |}
\hline
          & 1000 & 2000 & 5000  & 10~000 \\
\hline
~$ x $~ & $ x~y~z~(5) $ & $ x~y~z~(5) $ & $ x~y~z~(5) $ & $ x~y~z~(5) $ \\
\hline
~$ y $~ & $ x~y~(5) $ & $ x~y~(5) $ & $ x~y~(5) $ & $ x~y~(5) $ \\
\hline
~$ z $~ & $ z~(4),~y~z~(1)$ & $ z~(2),~x~z~(3) $ & $ x~z~(5) $ & $ x~z~(5) $ \\
\hline
\end{tabular}\\
\end{center}
    \label{tab:roessler_data_points}
\end{table}

We could confirm in the previous investigation that 
the correct relationships are obtained as the number of data points increases.
We next investigate the influence of observational noise
using four different noise levels.
We use 10~000 data points and the data are contaminated 
by Gaussian observational noise with the mean~zero 
and the standard deviation~0.01, 0.02, 0.05 and 0.1.
We prepare five sets of time series for each noise level.
Choosing a time delay up to 20 for time series of each data and 
the constant function give 61 candidate basis functions 
in the dictionary.
Using the dictionary we build the multivariate RAR models.
The summarized information are shown in Table~\ref{tab:roessler_noise}.
The information for $ x $ are correct in all cases,
and the correct information for $ y $ and $ z $ is obtained 
when the observational noise level is 0.01, 0.02 and 0.05.
However, when the observational noise level is 0.1,
although the correct information is obtained,
the wrong information is also obtained.
This indicates that there are cases that the correct information cannot be 
obtained when the observational noise level is large.
%
\begin{table}[!t]
\caption{Investigation of the influence of the observational noise.
         The components contained in the multivariate RAR models
         using data generated by the R{\"o}ssler systems shown 
         in Fig.~\ref{fig:roessler_data}.
         The number of data points is 10~000,
         five sets of time series are used,
         and the data are contaminated by Gaussian observational noise 
         with the mean~zero and the standard deviation~0.01, 0.02, 0.05 and 0.1.
         The value in parentheses is the number of times the result occurred.}
\begin{center}
\footnotesize
\begin{tabular}{| c | c | c | c | c | c |}
\hline
          & 0.01 & 0.02 & 0.05  & 0.1 \\
\hline
~$ x $~ & $ x~y~z~(5) $ & $ x~y~z~(5) $ & $ x~y~z~(5) $ & $ x~y~z~(5) $ \\
\hline
~$ y $~ & $ x~y~(5) $ & $ x~y~(5) $ & $ x~y~(5) $ & $ x~y~(2),~x~y~z~(3) $ \\
\hline
~$ z $~ & $ x~z~(5) $ & $ x~z~(5) $ & $ x~z~(5) $ & $ x~z~(1),~y~z~(3),~x~y~z~(1) $ \\
\hline
\end{tabular}\\
\end{center}
    \label{tab:roessler_noise}
\end{table}
\section{Reexamination of the proposed method}
\label{sec:reexamination_of_the_proposed_method}

In building an RAR model, it is necessary to select terms with important 
time delays with an appropriate information criterion to find the optimal model.
The obtained optimal model is thus influenced by the employed combination of 
the selection method and the information criterion.

A variety of information criteria, such as Akaike Information Criterion,
Schwarz Information Criterion, Description Length, and so on,
have already been proposed with their own different 
backgrounds~\cite{Nakamura-etal:IC2006}.
It means that the optimal models corresponding to different information criteria 
are not necessarily identical, even if we compare all possible values of 
the criteria calculating all possible combinations of terms.
As each best model reflects each background of the employed information criterion,
we should be careful in comparing the results by taking
these backgrounds into consideration.

Apart from the selection of information criterion,
the calculation of possible combinations of terms causes another concern.
The number of all combinations explodes as the number of components 
in the time series increases.
Selecting the optimal subset from a dictionary of basis functions 
thus becomes an NP-hard problem
and has to be solved heuristically~\cite{Judd-Mees:1995}.
Various heuristic algorithms have been proposed for selecting basis 
functions~\cite{Judd-Mees:1995,Nakamura-etal:2004,Tibshirani:losso1996}
and some optimization approaches can be applied for the modelling,
for example, simulated annealing, genetic algorithm, deep learning and 
so on~\cite{Vidal:SA_book1993,Holland:GA_book1992,Ohlsson:deep_learning_book2011}.
Also, another idea has been introduced to impose a sparseness constraint 
onto MVAR model~\cite{Davis-etal:MVAR_model}.
An important point when we employ non-exhaustive search is that
the selected model might be not optimal 
but nearly optimal corresponding to a local minimum
of an employed information criterion~\cite{Judd-Mees:1998}.
Hence, we should take care in choosing a selection algorithm
and should check the plausibility of the selected model.
In this paper, we choose the Description Length modified by Judd and Mees
for the information criterion~\cite{Judd-Mees:1995,Judd-Mees:1998} 
and the selection method using total error for the heuristic 
algorithm~\cite{Nakamura-etal:2004}.
The reason for this choice is simply
because this combination of information criterion and selection method
has been proven to be effective in modelling both linear and 
nonlinear dynamics and to obtain better models in many 
cases~\cite{Nakamura-etal:2004}.
Although we understand that there are many other alternatives,
exhaustive comparison between them would be far beyond the scope of this work.
However, we note that an idea of using summarized information of the RAR model
to construct the directed network is central to this approach.

The proposed method~(and the current approaches alike) does not work well
when there is no linear periodic structure in the data.
One typical example is the Logistic map~\cite{May:logistic_map1976}.
It is well known that the Logistic map is a nonlinear system that lacks 
clear periodicity, as the randomness is equivalent to that of 
IID random variables. 
The proposed method and the current approaches as well
cannot treat such a data appropriately.
We need an alternative approach to tackle it theoretically.
\subsection{Connection to Granger causality}
\label{sec:connection_to_granger_causality}

The proposed method constructs a directed network using 
components composed of multivariate RAR models.
We consider that the Granger causality using MVAR models provides 
a similar statistical approach~\cite{Granger:1969}.

We consider that the Granger causality is useful, but also very restrictive
because of the following two problems.
Although these problems will be mentioned separately, 
they are strongly interconnected. 
One is to use MVAR models, and the other is to use prediction accuracy.
As mentioned in Section~\ref{sec:problems_with_current_approaches},
MVAR models have difficulties in treating peculiarities of data appropriately.
Such a model often becomes unstable and also has difficulty in prediction.
We need to treat the prediction accuracy carefully~\cite{Judd-Mees:1995}.
One of the strong reasons is that data available to us are usually contaminated 
by observational noise to a varying degree and 
the true dynamics in a phenomenon is intertwined with observational noise 
in time series.
Hence, the high prediction accuracy means that the model provides 
similar behaviour to not that of the true phenomenon but that of the noisy data.
That is, the model should not be fitted to the data too 
closely~\cite{Judd-Mees:1995}.

We consider that the true concept~(or true intent) of the Granger causality 
principle is that the components are recognized to have causality, 
if the role and importance of the components for a model cannot be ignored.
RAR models include only terms that contribute significantly to the model, 
as assessed by an information criterion.
Although RAR models does not refer to the causal relationship,
we consider that the proposed method adheres faithfully to the concept of 
the Granger causality in this sense.\footnote{As MVAR models contain all components,
    it is difficult to know relationships among the components from the formulae.
    However, if only necessary terms are contained in models,
    we can directly know the relationships among the components.
    Hence, if we can obtain such a model,
    elaborate approaches such as the Granger causality, DTF and PDC 
    would not be necessary.
    We consider that the proposed method is a simple approach
    which can meet this requirement.}
\section{Applications}
\label{sec:application}

Based on the thorough arguments and the results of these computational studies,
we apply the proposed method to two experimental systems:
(i)~hourly meteorological time series in Kobe, Japan 
and (ii)~multichannel electroencephalography time series with 10~channels
(measured during resting state with eyes closed).
As shown in Figs.~\ref{fig:meteorological_data} and \ref{fig:EEG_data},
each of them exhibits irregular fluctuations.

The naive method remains the most commonly used approach, 
because of its conceptual and computational simplicity~\cite{Mantegna:1999,%
Yamasaki-etal:2008,Nagy-etal:pigeon_flocks10}.
Hence, we also show the networks constructed by the naive method
from the same data sets for comparison.\footnote{Although the comparison 
    with PDC and DTF for all real world data might be useful, 
    as it is beyond our purpose, we apply the naive method only.}

\subsection{Meteorological data in Kobe, Japan}
\label{subsec:meteorological_data}

\begin{figure}[!t]
\centering
(a)\includegraphics[width=6.0cm]{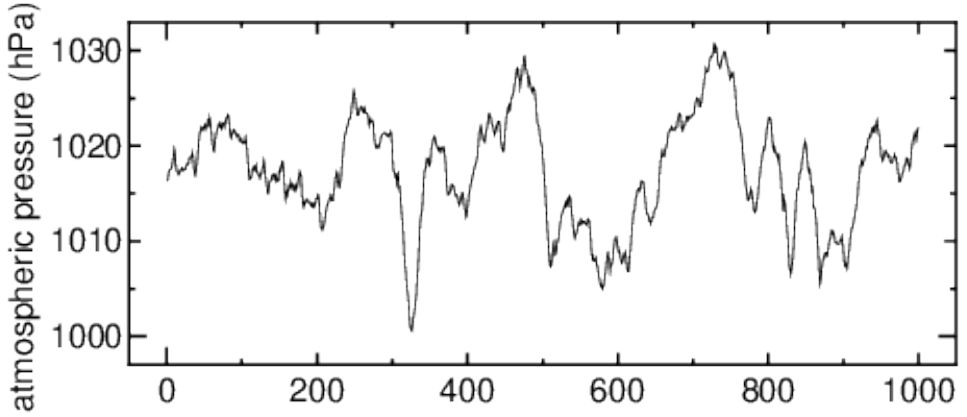}
~~(b)\includegraphics[width=6.0cm]{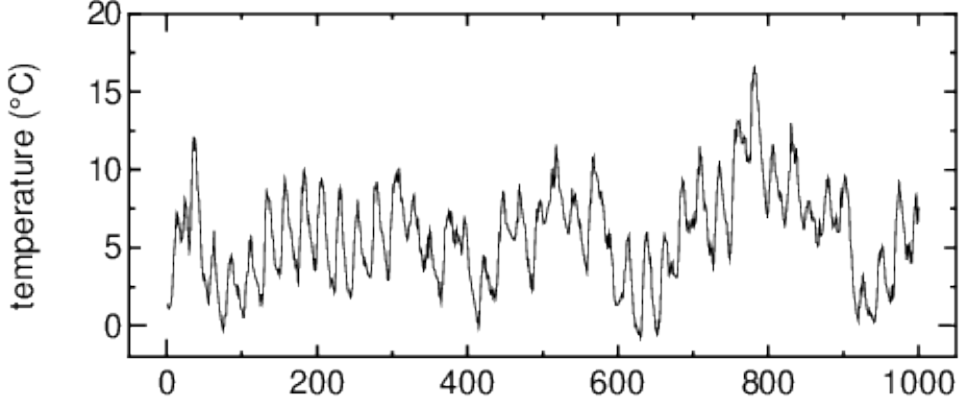}\\
(c)\includegraphics[width=6.0cm]{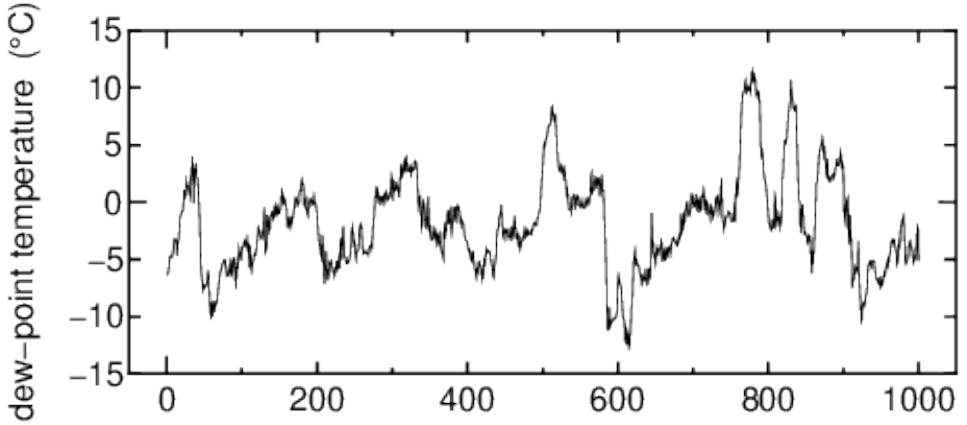}
~~(d)\includegraphics[width=6.0cm]{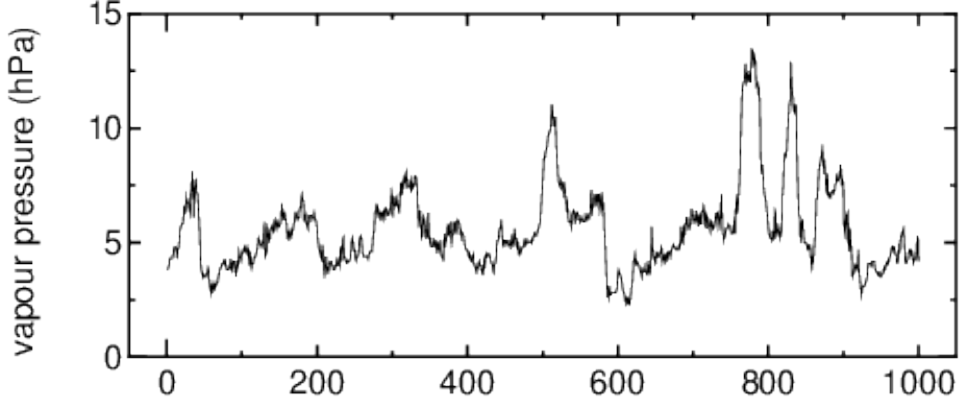}\\
(e)\includegraphics[width=6.0cm]{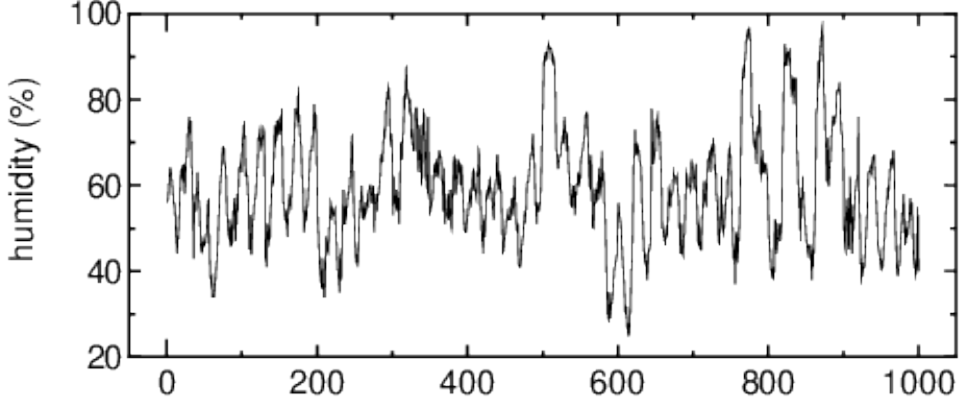}
\caption{Hourly meteorological time series in Kobe, Japan
          from 1 January to 12 February in 2013:
          (a)~atmospheric pressure,
          (b)~temperature,
          (c)~dew-point temperature,
          (d)~vapour pressure,
          and (e)~humidity.
          These data are used for building multivariate RAR models.}
\label{fig:meteorological_data}
\end{figure}

The meteorological data set consists of five time series:
the atmospheric pressure, the atmospheric temperature, the dew-point temperature, 
the vapour pressure and the humidity,
taken hourly in Kobe, Japan from 1 January to 12 February 
in 2013.\footnote{The data can be obtained from Japan Meteorological Agency,
                  \text{http://www.jma.go.jp/jma/indexe.html}}
The measurement location is 34$^\circ$--41.8$^{''}$ north latitude and 
135$^\circ$--12.7$^{''}$ east longitude.
From the profiles of the time series shown in Fig.~\ref{fig:meteorological_data},
the relationship among these five time series is complicated and hard to 
be extracted.

We use 1000~data points~(about 42 days) for building multivariate RAR models.
As there are five time series, choosing a time delay up to 15 for time series 
of each data and the constant function give 76 candidate basis functions 
in the dictionary.
Using the dictionary we build the multivariate RAR model for each data.
The summarized information of the obtained five multivariate RAR models are
\begin{align}
 x_1 &= f_{1}(x_2),               \label{eq:meteor_RAR_x1} \\
 x_2 &= f_{2}(x_4),               \label{eq:meteor_RAR_x2} \\
 x_3 &= 0,                        \label{eq:meteor_RAR_x3} \\
 x_4 &= f_{4}(x_3),               \label{eq:meteor_RAR_x4} \\
 x_5 &= f_{5}(x_2, x_3),          \label{eq:meteor_RAR_x5} 
\end{align}
where $ x_1 $ corresponds to the atmospheric pressure, 
$ x_2 $ the atmospheric temperature,
$ x_3 $ the dew-point temperature, 
$ x_4 $ the vapour pressure,
$ x_5 $ the humidity,
and zero means that there is no connection.

\begin{figure}[!t]
\centering
(a)\includegraphics[width=6.0cm]{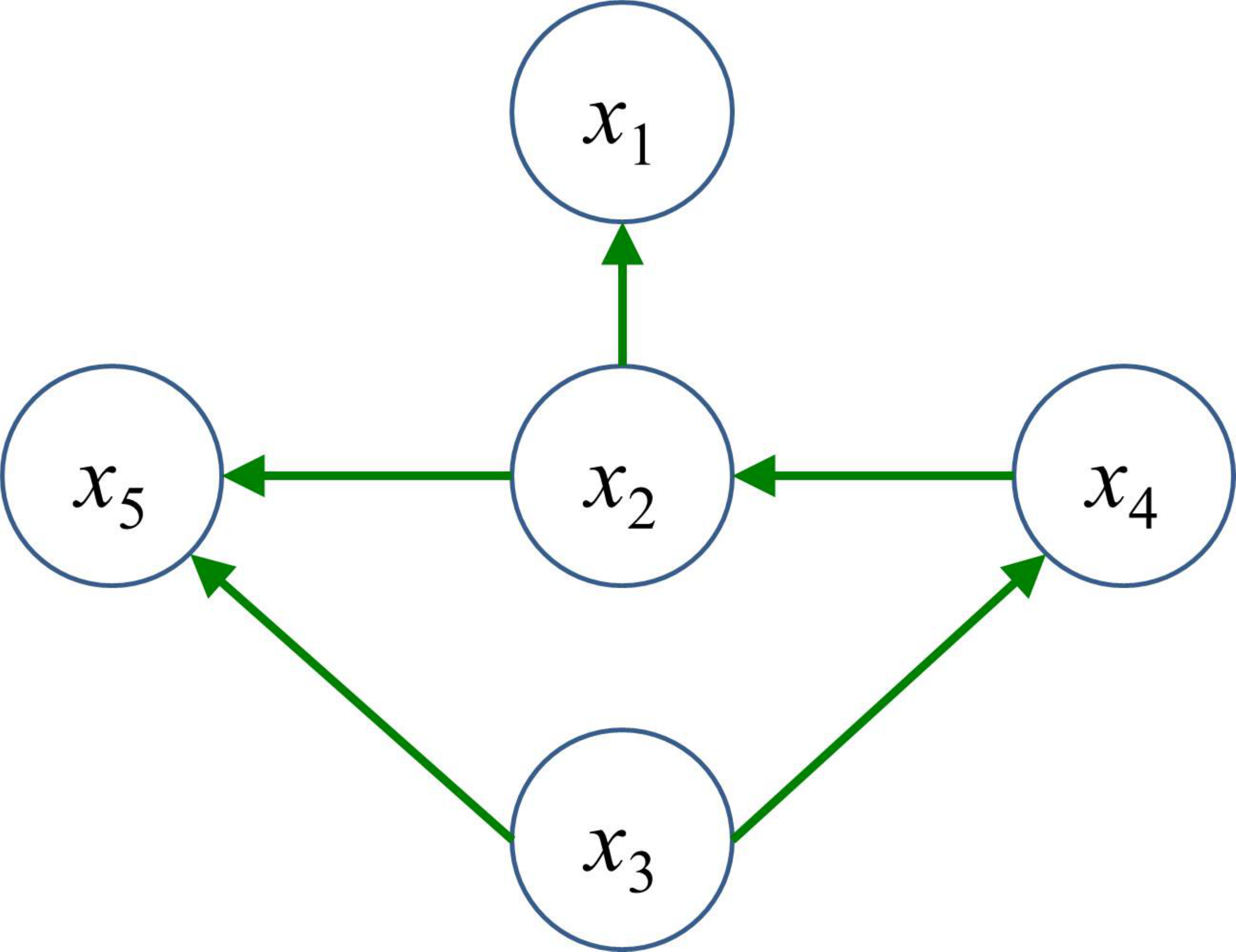}
~~(b)\includegraphics[width=6.0cm]{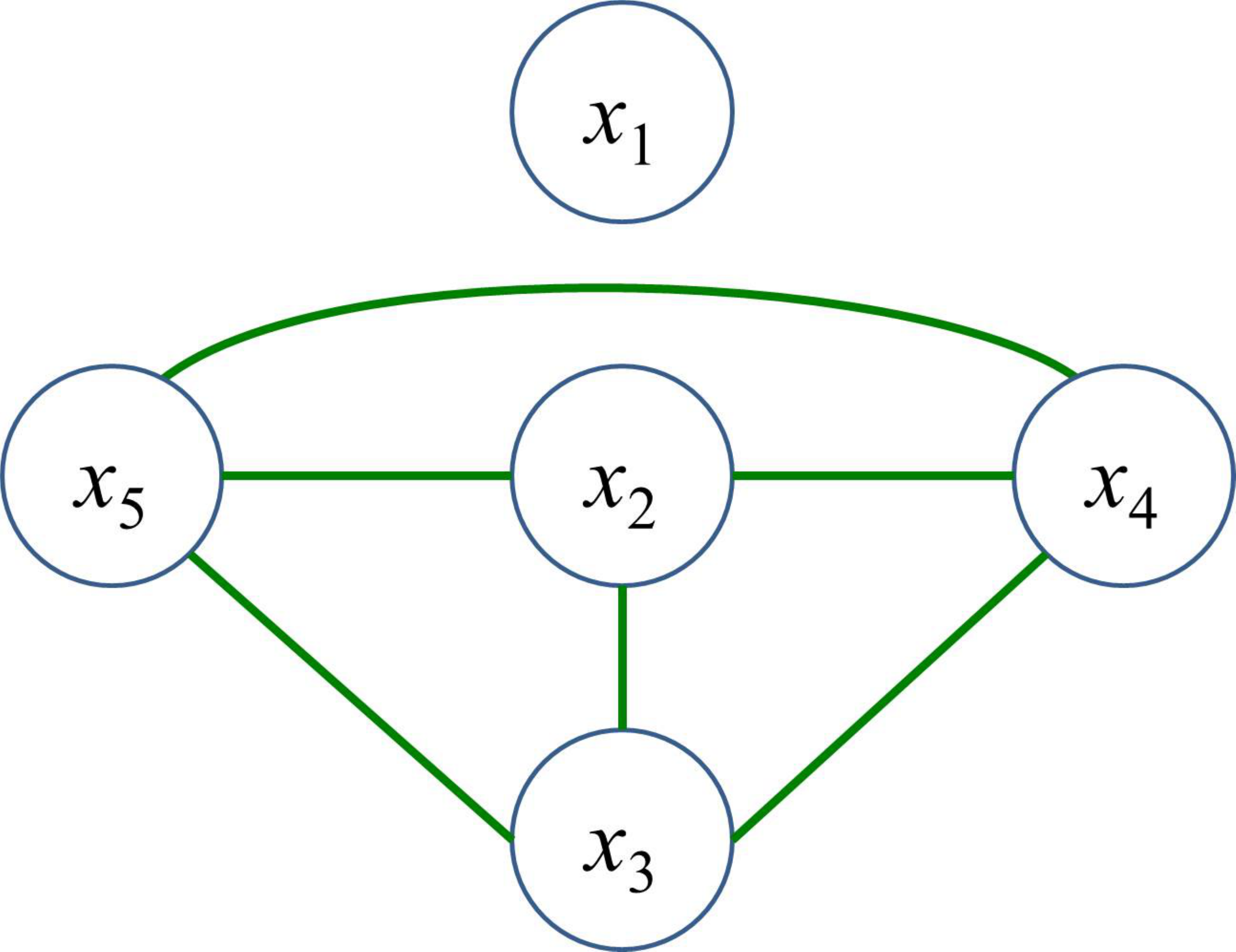}
\caption{(Colour online)~Networks of meteorological time series taken hourly 
         in Kobe, Japan from 1 January to 12 February in 2013:
         (a)~the directed network constructed by the proposed method,
         and (b)~the network constructed by the naive method using the CC 
         with the threshold~$ 0.5 $,
         where $ x_1 $ corresponds to atmospheric pressure, 
         $ x_2 $ temperature,
         $ x_3 $ dew-point temperature,
         $ x_4 $ vapour pressure,
         and $ x_5 $ humidity.}
\label{fig:nw_meteor}
\end{figure}

Figure~\ref{fig:nw_meteor}(a) shows the directed network constructed by 
the proposed method.
The numbers of in-degree and out-degree for each node in 
Fig.~\ref{fig:nw_meteor}(a) are shown in 
Table~\ref{tab:number_of_in_and_out-degree_meteor}.
From these results we find that the atmospheric pressure and the humidity 
are influenced by others but do not have influence on any other components.
On the contrary, the dew-point temperature is not influenced by others
but has influence on the other two components.
The atmospheric temperature and the vapour pressure are influenced by others 
and have influence on others at the same time.
We also found that there is no mutual~(bi-directional) connections 
among any component.

For comparison we show the network obtained by the naive method in 
Fig.~\ref{fig:nw_meteor}(b), where the cross correlation~(CC) function 
is evaluated between the time lag $ -15 $ and $ 15 $ with the threshold~$ 0.5 $.
Table~\ref{tab:CC_of_meteorological_data} shows all the values.
There are similarities and differences between Figs.~\ref{fig:nw_meteor}(a) 
and \ref{fig:nw_meteor}(b).
The link between $ x_1 $ and $ x_2 $ in Fig.~\ref{fig:nw_meteor}(a)
is absent in Fig.~\ref{fig:nw_meteor}(b).
The time dependencies of \(x_1\) and \(x_2\) in
Figs.~\ref{fig:meteorological_data}(a) and~\ref{fig:meteorological_data}(b),
are clearly dissimilar and the largest absolute value of the CC function is 
no more than $ 0.1779 $.
We consider that the RAR modelling technique uncovered a hidden relationship 
between these components.
On the other hand, there are two links between $ x_2 $ and $ x_3 $ and 
between $ x_4 $ and $ x_5 $ in Fig.~\ref{fig:nw_meteor}(b),
which are absent in Fig.~\ref{fig:nw_meteor}(a).
We consider that these ``redundant'' links can be understood
as ``indirect'' relationships deducible from the directed network 
in Fig.~\ref{fig:nw_meteor}(a). 
For example, the link between $ x_2 $ and $ x_3 $ in Fig.~\ref{fig:nw_meteor}(b) 
can be deduced from two consecutive directed links from $ x_3 $ to $ x_4 $ and 
from $ x_4 $ to $ x_2 $ in Fig.~\ref{fig:nw_meteor}(a).
The link between $ x_4 $ and $ x_5 $ in Fig.~\ref{fig:nw_meteor}(b) can also 
be deduced from two consecutive directed links from \(x_4\) to \(x_2\) and 
from \(x_2\) to \(x_5\).

We consider the interactions between these five~physical quantities.
As the air becomes lighter~(heavier) when the temperature becomes higher~(lower),
it is generally considered that the change of the temperature brings about 
the change of the atmospheric pressure.
As there is a directed link from $ x_2 $ to $ x_1 $ on the network 
as shown in Fig.~\ref{fig:nw_meteor}(a),
where $ x_1 $ is atmospheric pressure and $ x_2 $ is temperature,
we consider that the proposed method can correctly identify the relationship.
However, Fig.~\ref{fig:nw_meteor}(b) shows that there is no link between 
$ x_1 $ and $ x_2 $.
Hence, this indicates that the naive method fails to detect this relationship.
It is also known that there is no relationship between the temperature
and the dew-point temperature.
Although the proposed method shows that there is no relationship between 
the temperature and the dew-point temperature~($ x_2 $ and $ x_3 $),
the naive method shows that there is relationship between them.
%
\begin{table}[!t]
\caption{The number of in-degree and out-degree of the directed network
         of meteorological data shown in Fig.~\ref{fig:nw_meteor}(a).}
\begin{center}
\footnotesize
\begin{tabular}{| c | c | c | c | c | c |}
\hline
             & atmospheric & temperature & dew-point   & vapour    & humidity \\
             & pressure    &             & temperature & pressure &          \\
\hline
~in-degree~  &     1    &      1   &      0   &      1    &   2  \\
\hline
~out-degree~ &     0    &      2   &      2   &      1   &    0   \\
\hline
\end{tabular}
\end{center}
    \label{tab:number_of_in_and_out-degree_meteor}
\end{table}
%
\begin{table}[!t]
\caption{The largest absolute values of the CC function 
         of all possible pairs between the time lag $ -15 $ and $ 15 $,
         where the number in the parentheses is the time lag 
         when the CC function has the largest absolute value.
         The data are hourly meteorological time series shown 
         in Fig.~\ref{fig:meteorological_data}.
         The values of the CC function are estimated using 
         1000~data points.}
\label{tab:CC_of_meteorological_data}
\begin{center}
\footnotesize
\begin{tabular}{|c|c|c|c|c|c|c|}
\hline
   & $ x_1 $ & $ x_2 $ & $ x_3 $ & $ x_4 $ & $ x_5 $\\
\hline
$ ~~~~x_1~~~~ $ & 1.0000 & ---  & --- & --- & --- \\
\hline
    $ x_2 $  & 0.1779 (-3) & 1.0000 & --- & --- & --- \\
\hline
    $ x_3 $  & 0.2622 (-6) & 0.6977 (-3) & 1.0000 & --- & --- \\
\hline
    $ x_4 $  & 0.2996 (-4) & 0.6951 (-2) & 0.9778 (0) & 1.0000 & --- \\
\hline
    $ x_5 $  & 0.2667 (-8) & 0.5634 (-11) & 0.7341 (0) & 0.7183 (0) & 1.0000 \\
\hline
\end{tabular}
\end{center}
\end{table}

We can explore the obtained network using the summarized information,
Eqs.~(\ref{eq:meteor_RAR_x1})--(\ref{eq:meteor_RAR_x5}).
We expect that this approach is effective for more complicated cases.
We consider the case between $ x_2 $ and $ x_3 $ again.
Eq.~(\ref{eq:meteor_RAR_x2}) is $ x_2 = f_{2}(x_4) $
and Eq.~(\ref{eq:meteor_RAR_x4}) is $ x_4 = f_{4}(x_3) $.
As Eq.~(\ref{eq:meteor_RAR_x2}) shows that $ x_4 $ is included in $ f_{2} $,
we can rearrange Eq.~(\ref{eq:meteor_RAR_x2}) using Eq.~(\ref{eq:meteor_RAR_x4}).
The rearrangement gives
\begin{align}
 x_2 &= f_{2}(x_4),          \label{eq:rearragement_x2-1} \\
     &= f_{2}(f_{4}(x_3)),   \label{eq:rearragement_x2-2} \\
     &= f'_{2}(x_3),         \label{eq:rearragement_x2-3} 
\end{align}
where $ f'_{2} $ is a new function for $ x_2 $.
Eq.~(\ref{eq:rearragement_x2-3}) explicitly shows
the ``indirect'' relationship between 
$ x_2 $ and~$ x_3 $.

\subsection{Electroencephalogram~(EEG) data}
\label{subsec:eeg_data}

\begin{figure}[!t]
\centering
  \includegraphics[width=4.0cm]{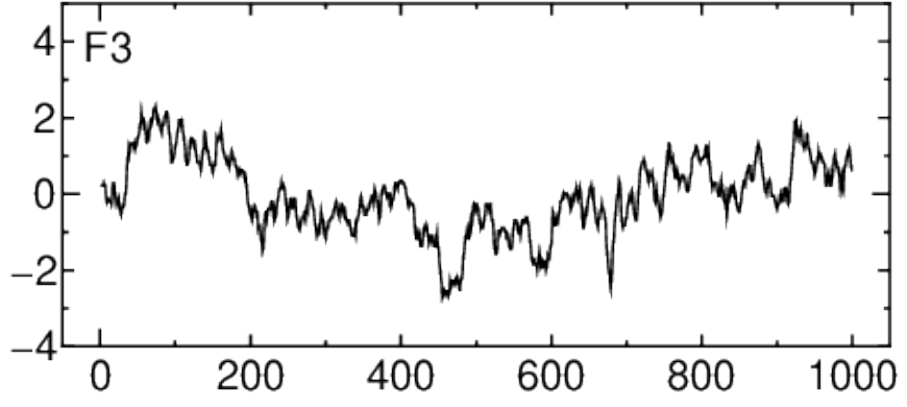}
~~\includegraphics[width=4.0cm]{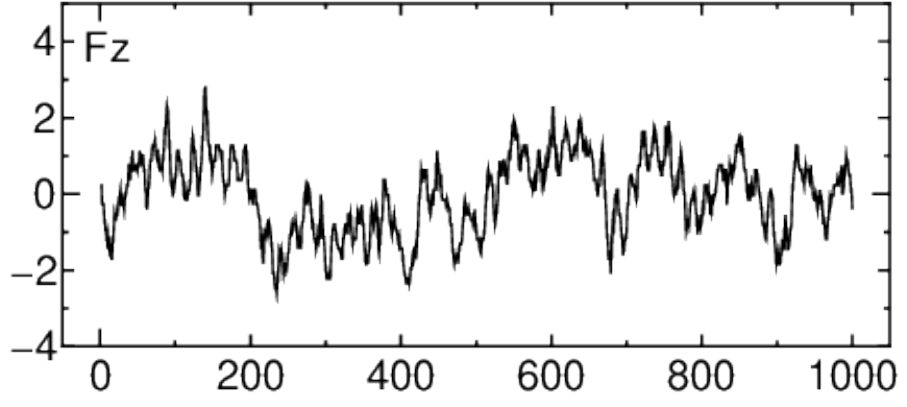}
~~\includegraphics[width=4.0cm]{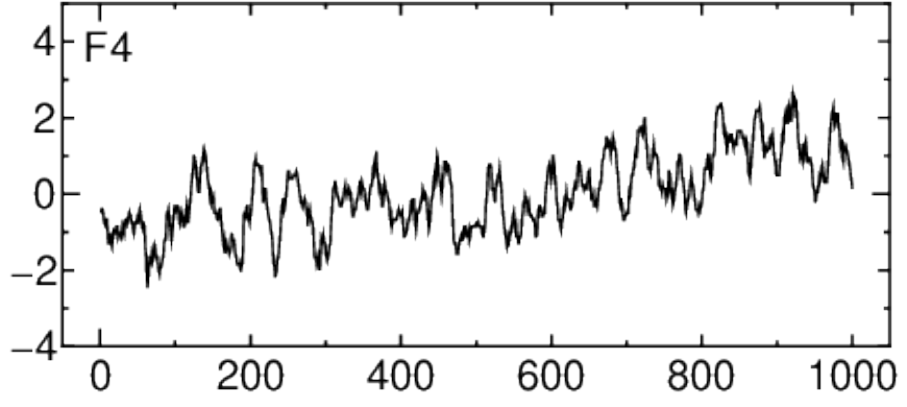}\\
  \includegraphics[width=4.0cm]{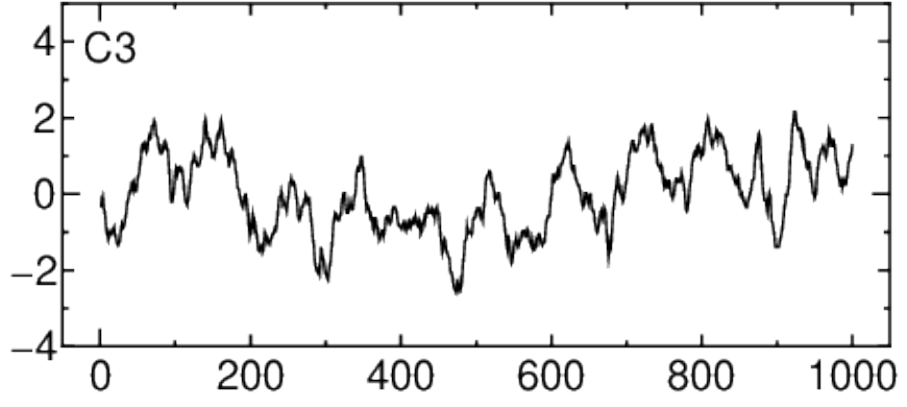}
~~\includegraphics[width=4.0cm]{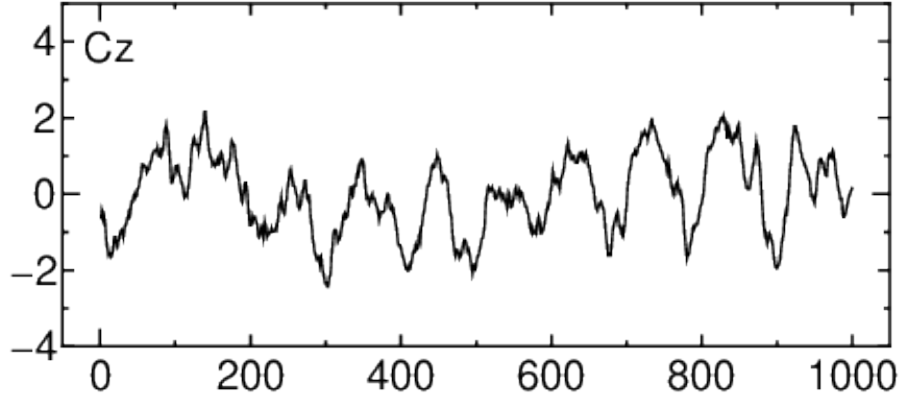}
~~\includegraphics[width=4.0cm]{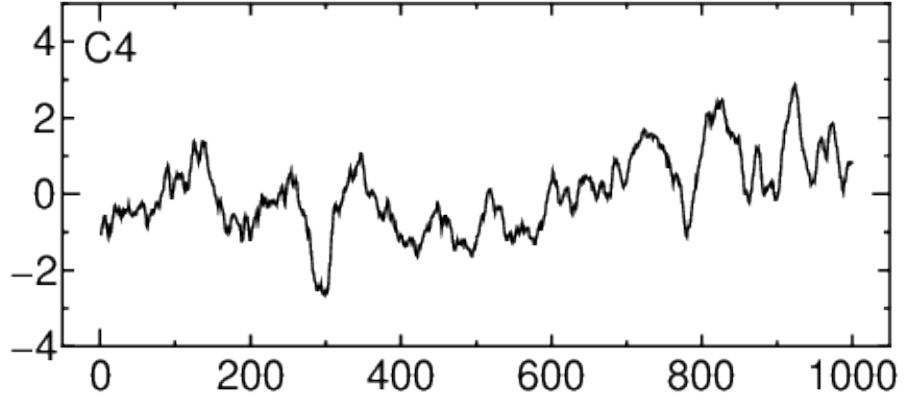}\\
  \includegraphics[width=4.0cm]{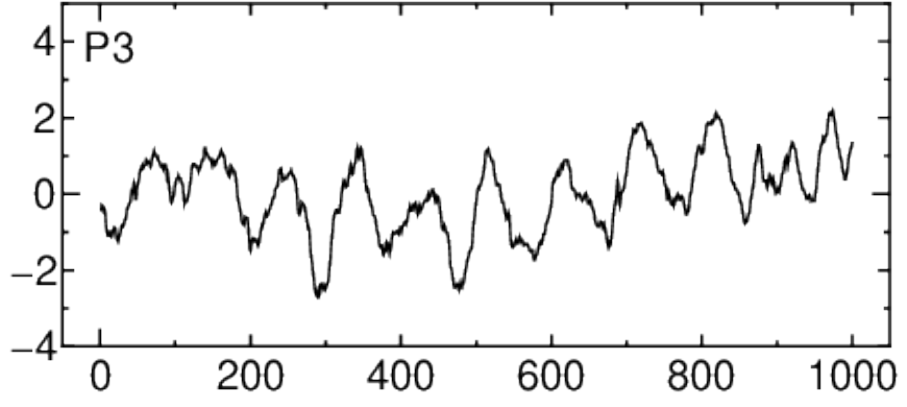}
~~\includegraphics[width=4.0cm]{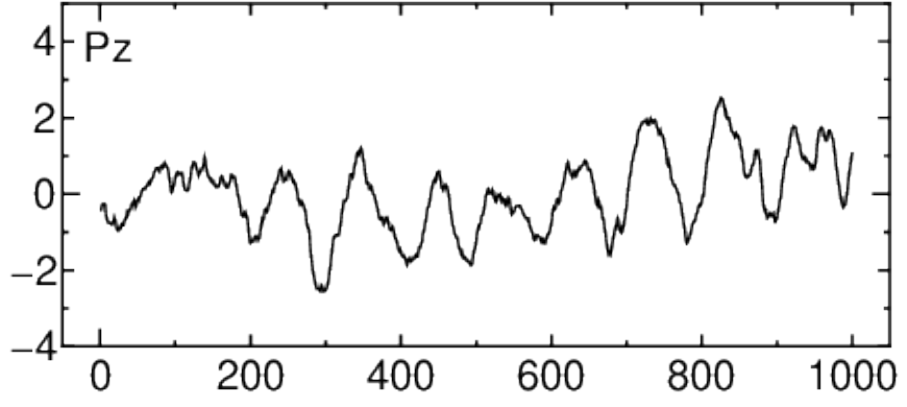}
~~\includegraphics[width=4.0cm]{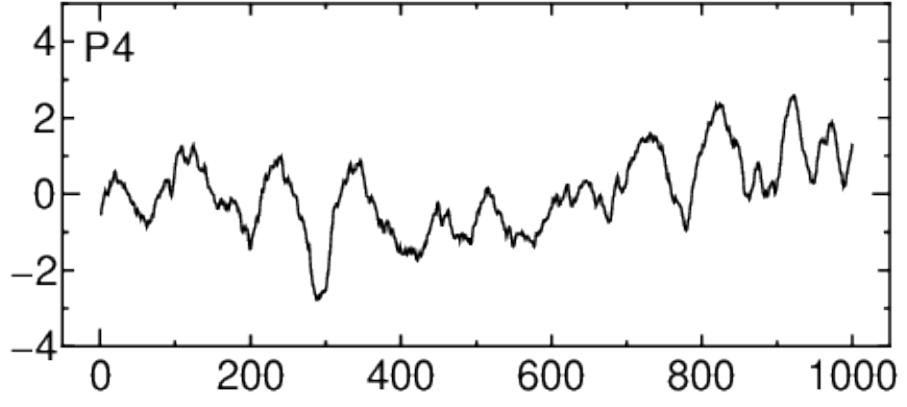}\\
  \includegraphics[width=4.0cm]{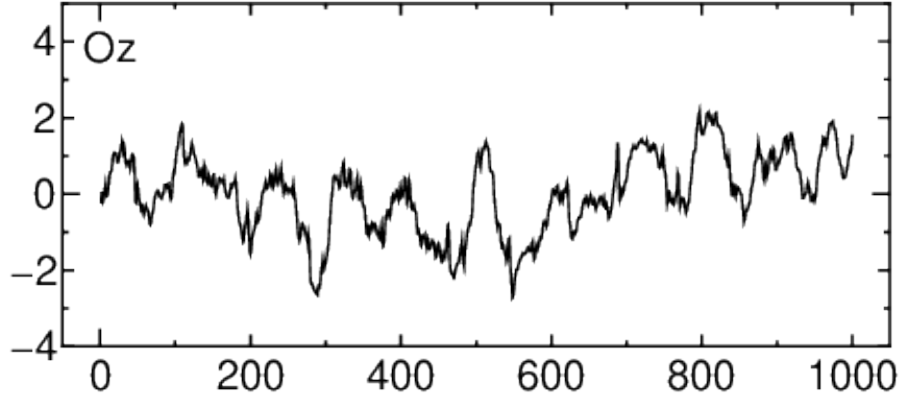}
\caption{Multichannel electroencephalography~(EEG) time series
          of Fz, Cz, Pz, Oz, F3, F4, C3, C4, P3 and P4 used 
          for building multivariate RAR models.}
\label{fig:EEG_data}
\end{figure}

The second application is to EEG data.
The EEG signal we use here was recorded from a healthy human adult 
during resting state with eyes closed in an electrically shielded room.
The EEG data were simultaneously obtained from 10 channels of 
Fz, Cz, Pz, Oz, F3, F4, C3, C4, P3 and P4 of the unipolar 10-20 Jasper 
registration scheme~\cite{Jasper:10-20}.
Mono-polar recordings, referenced to linked earlobes, were obtained 
from these channels using an Electrocap.
Vertical and horizontal eye movements were recorded, respectively, 
from electrode sites above and below the right eye and from near 
the outer canthi of each eye.  
Artefact corrupted records were removed from the analyses.  
The data were digitized at 1024~Hz using a twelve-bit digitizer. 
The EEG impedances were less than 5[K$ \Omega $]. 
The data were amplified by gain = 18~000, and amplifier frequency
cut-off settings of 0.03~Hz and 200~Hz were used~\cite{Rapp-etal:2005}.
We use 1000 data points~(around 1 second) to build multivariate RAR models.
Figure~\ref{fig:EEG_data} shows time series of each channel.

As there are 10~channels, choosing a time delay up to 25 for time series 
of each channel and the constant function give 251 candidate basis functions 
in the dictionary.
Using the dictionary we build the multivariate RAR model for each channel.
The summarized information of the obtained 10 multivariate RAR models are
\begin{align}
 Fz &= f_{Fz}(Cz),                      \label{eq:EEG_RAR_Fz} \\
 Cz &= f_{Cz}(Pz, Oz, F3),              \label{eq:EEG_RAR_Cz} \\
 Pz &= f_{Pz}(Fz, Cz, Oz, C3, P3, P4),  \label{eq:EEG_RAR_Pz} \\
 Oz &= f_{Oz}(Fz, Cz),                  \label{eq:EEG_RAR_Oz} \\
 F3 &= 0,                               \label{eq:EEG_RAR_F3} \\
 F4 &= f_{F4}(Fz),                      \label{eq:EEG_RAR_F4} \\
 C3 &= f_{C3}(F3, P3),                  \label{eq:EEG_RAR_C3} \\
 C4 &= f_{C4}(Fz, P4),                  \label{eq:EEG_RAR_C4} \\
 P3 &= f_{P3}(Pz, Oz),                  \label{eq:EEG_RAR_P3} \\
 P4 &= f_{P4}(Fz, Cz, Pz),              \label{eq:EEG_RAR_P4}
\end{align}
where zero means that there is no connection.

\begin{figure}[!t]
\centering
(a)\includegraphics[width=5.7cm]{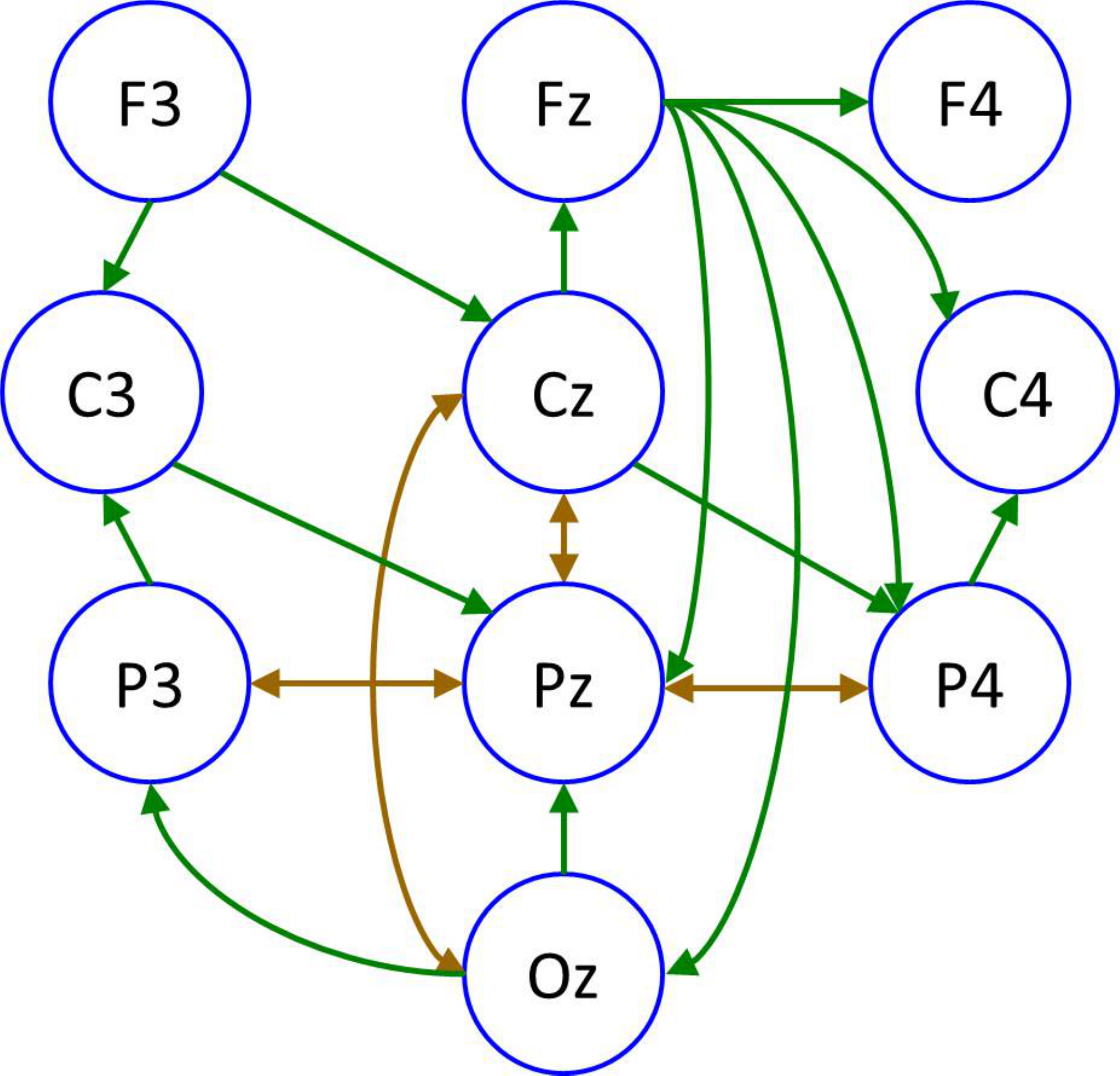}
~~(b)\includegraphics[width=6.0cm]{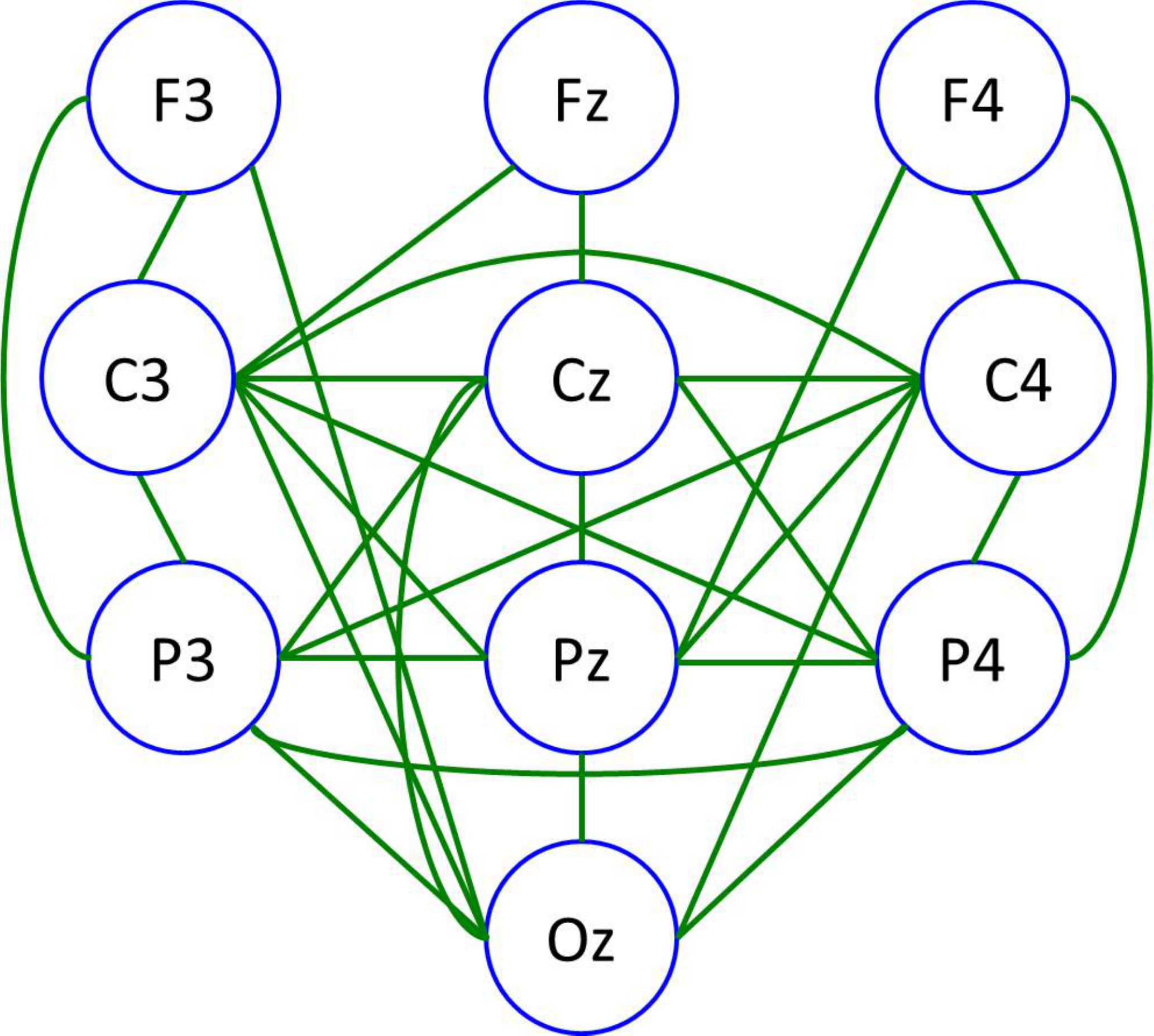}
 \caption{(Colour online)~Network of EEG data shown in Fig.~\ref{fig:EEG_data}:
          (a)~the directed network constructed by the proposed method,
          and (b)~the network constructed by the naive method using the CC 
          with the threshold~$ 0.5 $.}
\label{fig:nw_EEG}
\end{figure}

Figure~\ref{fig:nw_EEG} shows the network obtained by the proposed method
and the one obtained by the naive method.
In Fig.~\ref{fig:nw_EEG}(a) obtained by the proposed method, 
the channels, Cz, Pz, P3, P4, and Oz, are mutually connected by bi-directional 
arrows.
Figure~\ref{fig:nw_EEG}(b) shows that the network obtained by the naive method
has clearly more links than that by the proposed method.
This result might indicate that the constructed network by the naive method
has redundant links.
%
\begin{table}[!t]
\caption{The numbers of in-degree and out-degree for each node of 
         the directed network of EEG data shown in Fig.~\ref{fig:nw_EEG}(a).}
\label{fig:deg_sized_nws}
\begin{center}
\footnotesize
\begin{tabular}{|c|c c c c c c c c c c|}
\hline
             & ~Fz & ~Cz & ~Pz & ~Oz & ~F3 & ~F4 & ~C3 & ~C4 & ~P3 & ~P4~~\\
\hline
~in-degree~  &   1  &  3  &  6 &   2 &   0 &   1 &   2 &   2 &   2 &   3 \\
\hline
~out-degree~ &   5  &  4  &  3 &   3 &   2 &   0 &   1 &   0 &   2 &   2 \\
\hline
\end{tabular}
\end{center}
\label{tab:number_of_in_and_out-degree_EEG}
\end{table}
The numbers of in-degree and out-degree for each node in Fig.~\ref{fig:nw_EEG}(a) 
are shown in Table~\ref{tab:number_of_in_and_out-degree_EEG}. 
Table~\ref{tab:number_of_in_and_out-degree_EEG} shows 
that the node~Pz has the largest in-degree and the node~Fz has the largest 
out-degree.
The largest in-degree of node~Pz implies that the dynamics of the parietal areas 
captured by node~Pz is under the influences of many other brain areas distributed 
over parietal and occipital parts of the brain.
Since the parietal association areas are considered to be the parts that 
integrates sensory information from widely distributed parts of the brain,
this result seems plausible.
On the other hand, the largest out-degree of node~Fz implies that 
the frontal areas recorded with node~Fz has the influence over a wide-range 
of the brain. 
This is natural as the frontal brain areas are regarded to be involved 
in cognitive control and project top-down connections to lower brain areas.

We finally consider effects of the surface Laplacian briefly.
Functional connectivity estimation with the proposed method may be
contaminated by volume conduction effects~\cite{Srinivasan-etal:2007}.
Such effect would be relatively small for the current result 
as the RAR models utilizes terms with time delays,
not just zero-lag correlations, 
to estimates the connections between nodes, 
although the RAR based on scalp EEG may not be free of the volume conduction 
as pointed out for DTF~\cite{Brunner-etal:2016,Steen-etal:2016,%
Kaminski-Blinowska:2017}. 
Also, the number of electrodes used in this study is just 10, 
implying that the volume conduction effect was small. 
For high density EEG where the volume conduction effect is more serious, 
the methods such as surface Laplacian~\cite{Srinivasan-etal:2007}
or source localization techniques would be useful~\cite{Haufe-etal:2013}. 
Further studies would be necessary to investigate the effects of 
the volume conduction on the RAR based methods for functional 
connectivity estimation in a sensor space.

\section{Summary}
\label{sec:summary}

We have described an algorithm from constructing directed networks
from multivariate time series based on the RAR model
from the perspective of linear periodic structures~(relationships).
We also described the theoretical problems with the current approaches.
We showed that the proposed method can extract reliable linear periodic 
structures among time series and that the constructed networks are 
considered to be faithful representation of the dynamical relationships.
Our arguments and computational results show that 
the proposed method rectifies the drawbacks contained in the current approaches
and constructs networks from multivariate time series faithfully.

We note that there are restrictions when applying the proposed method.
RAR model cannot always identify nonlinear periodic structure in the data,
and RAR model built using selection algorithms might not be the optimal.
Hence, the network constructed by the proposed method might be 
near-optimal network in some cases.
Moreover, it is difficult to build appropriate RAR models
when the observational noise is large.
Despite of these concerns,
we believe that our method and the constructed networks has
a wide range of applicability to real-world phenomena
and provide us with useful information by using with care.
\section{Acknowledgement}

The authors thank to the anonymous referees for various valuable comments
and suggestions.
T.T. would like to acknowledge the support of 
a Grant-in-Aid for Scientific Research~(C) (No.~24540419) 
from the Japan Society for the Promotion of Science~(JSPS).
M.S. is funded by Discovery Project~(DP140100203).
T.N. and the other authors would like to acknowledge Professor Paul~E.~Rapp 
(Uniformed Services University of the Health Sciences) 
for providing us with the EEG data used in Ref~\cite{Rapp-etal:2005}.
\appendix
\section{Detecting periodicities from linear and nonlinear data using RAR model}
\label{sec:detection_by_RAR_model}

\begin{figure}[!t]
\centering
(a)\includegraphics[width=6.0cm]{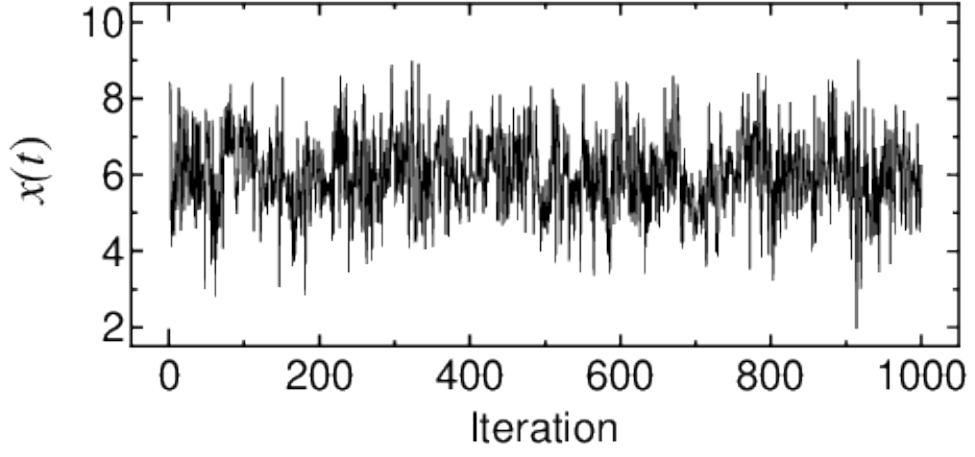}
~~(b)\includegraphics[width=6.0cm]{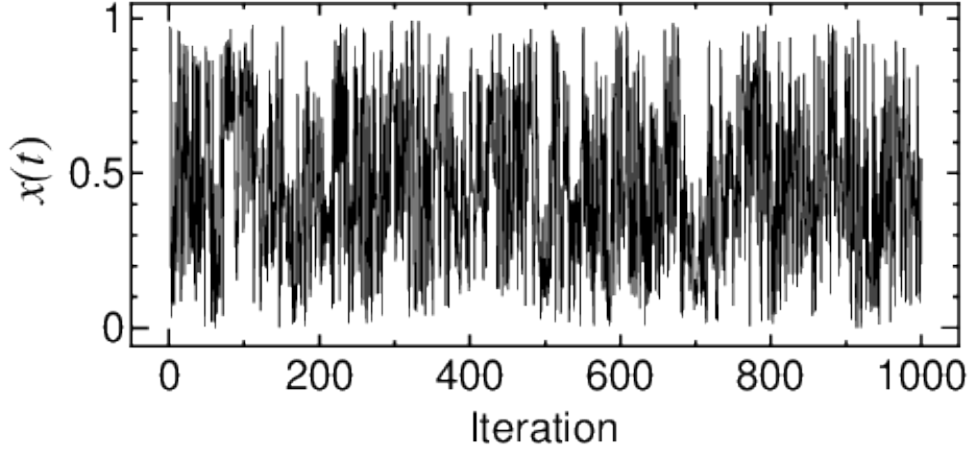}\\
(c)\includegraphics[width=6.0cm]{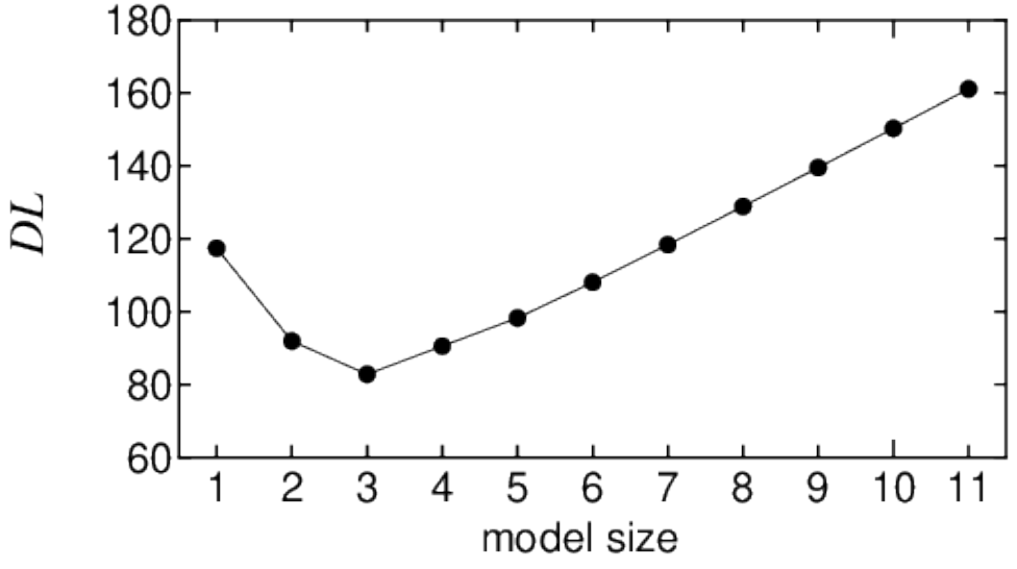}
~~(d)\includegraphics[width=6.0cm]{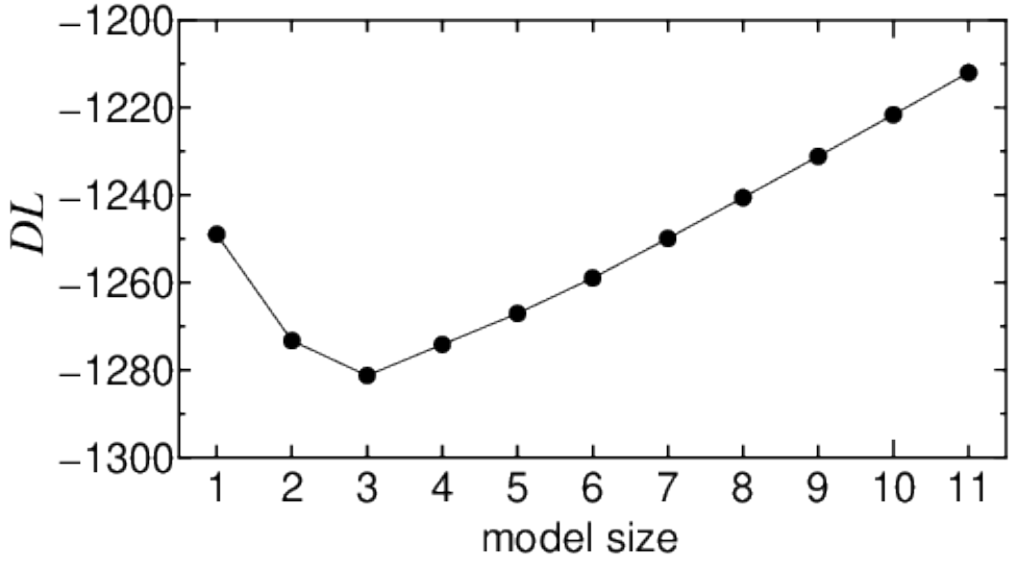}
\caption{The behaviours of linear and nonlinear data
         and plots of description length~$ DL $ against size of model 
         using the bottom-up method using the total error.
         We use these data to build RAR models to identify periodicities 
         in the data.
         (a)~linear data from Eq.~(\ref{eq:linear_RAR}),
         (b)~nonlinear data transformed by a nonlinear function, 
             Eq.~(\ref{eq:nonlinear_filter}),
         (c)~description length and model size of linear data shown 
             in Fig.~\ref{fig:linear&nonlinear_RAR_data}(a),
         and (d)~description length and model size of linear data shown 
         in Fig.~\ref{fig:linear&nonlinear_RAR_data}(b),}
\label{fig:linear&nonlinear_RAR_data}
\end{figure}

We show that RAR model can identify periodicities in data,
irrespective of whether the data is linear or nonlinear.
We use the following system to generate linear data:
\begin{eqnarray}
    x(t) = a_0 + a_1 \; x(t-1) + a_6 \; x(t-6) + \eta(t),
    \label{eq:linear_RAR}
\end{eqnarray}
where $ a_0 = 2.945~206 $, $ a_1 = 0.300~739 $, $ a_6 = 0.202~056 $,
and $ \eta(t) $ is dynamic noise, IID Gaussian random variables 
with mean zero and standard deviation~1.0~\cite{Small-Judd:1999}.
The behaviour of the time series is shown in Fig.~\ref{fig:linear&nonlinear_RAR_data}(a).

For periodicity detection in nonlinear data,
we distort the data generated by Eq.~(\ref{eq:linear_RAR}) by
a static monotonic nonlinear function~$ h(x) $, 
\begin{eqnarray}
h(x) = \frac{\left[ \frac{x - x_{min} - 0.0001}{x_{max}-x + 0.0001} \right]^\rho}
       {1+\left[ \frac{x - x_{min} - 0.0001}{x_{max}-x + 0.0001} \right]^\rho},
    \label{eq:nonlinear_filter}
\end{eqnarray}
where $ x_{min} $ and $ x_{max} $ are the minimum and maximum value of $ x(t) $ 
in the original time series, and $ \rho=3 $~\cite{Rapp-etal:1993}.
The behaviour of the time series is shown in Fig.~\ref{fig:linear&nonlinear_RAR_data}(b).
It should be noticed that the nonlinear data have the same periodicities 
as those of the linear data, even though the nonlinear data are heavily distorted 
by a nonlinear function.

As the observational data, we use the 1000~data points with Gaussian observational
noise with the mean zero and the standard deviation~0.01 for linear data and
0.003 for nonlinear data, respectively.
These noise levels are equivalent to $ 1 \% $ to both the data.
Choosing a time delay=10 gives 11 candidate basis functions in the dictionary. 
These are the constant function and the linear terms, 
$ x(1-1), x(t-2), \dots, x(t-10) $. 
We apply the bottom-up method using the total error and the exhaustive search 
both to the dictionary of the linear data and that of the nonlinear data.
As shown in Figs.~\ref{fig:linear&nonlinear_RAR_data}(c) and 
\ref{fig:linear&nonlinear_RAR_data}(d),
the description length is the smallest when the model size is $ 3 $ 
in both the linear and the nonlinear cases.
In both cases, only the terms included in Eq.~(\ref{eq:linear_RAR}) 
(a constant, $ x(t-1) $, and $ x(t-6) $) are selected.
That is, the correct model is selected as the best model.
The exhaustive search also selects the correct model as the best model.
This result indicates that the RAR model is effective in identifying 
periodicities for linear and nonlinear data.


\begin{thebibliography}{10}
\expandafter\ifx\csname url\endcsname\relax
  \def\url#1{\texttt{#1}}\fi
\expandafter\ifx\csname urlprefix\endcsname\relax\def\urlprefix{URL }\fi
\expandafter\ifx\csname href\endcsname\relax
  \def\href#1#2{#2} \def\path#1{#1}\fi

\bibitem{Ohira-Yamane:2000}
T.~Ohira, T.~Yamane, Delayed stochastic systems, Physical Review E 61 (2000)
  1247--1257.

\bibitem{Kaminski-Blinowska:DTF1991}
M.~J. Kami\'{n}ski, K.~J. Blinowska, A new method of the description of the
  information flow in brain structures, Biological Cybernetics 65 (1991)
  203--210.

\bibitem{Kaminski-etal:DTF2001}
M.~Kami\'{n}ski, M.~Ding, W.~A. Truccolo, S.~L. Bressler, Evaluating causal
  relations in neural systems: granger causality, directed transfer function
  and statistical assessment of significance, Biological Cybernetics 85 (2001)
  145--157.

\bibitem{Baccala-Sameshima:PDF2001}
L.~A. Baccal\'{a}, K.~Sameshima, Partial directed coherence: a new concept in
  neural structure determination, Biological Cybernetics 84 (2001) 463--474.

\bibitem{Mantegna:1999}
R.~N. Mantegna, Hierarchical structure in financial markets, The European
  Physical Journal B 11~(1) (1999) 193--197.

\bibitem{Farkas-etal:2003}
I.~Farkas, H.~Jeong, T.~Vicsek, A.-L. Barab\'{a}si, Z.~N. Oltvai, The topology
  of the transcription regulatory network in the yeast, saccharomyces
  cerevisiae, Physica A 318~(3--4) (2003) 601--612.

\bibitem{Astolfi-etal:2007}
L.~Astolfi, F.~Cincotti, D.~Mattia, M.~G. Marciani, L.~A. Baccala,
  F.~de~Vico~Fallani, S.~Salinari, M.~Ursino, M.~Zavaglia, L.~Ding, J.~C.
  Edgar, G.~A. Miller, B.~He, F.~Babiloni, Comparison of different cortical
  connectivity estimators for high-resolution eeg recordings, Human brain
  mapping 28~(2) (2007) 143--157.

\bibitem{Yamasaki-etal:2008}
K.~Yamasaki, A.~Gozolchiani, S.~Havlin, Climate networks around the globe are
  significantly affected by el ni{\~n}o, Physical Review Letters 100~(22)
  (2008) 228501.

\bibitem{Tsonis-Swanson:2008}
A.~Tsonis, K.~Swanson, Topology and predictability of el ni{\~n}o and la
  ni{\~n}a networks, Physical Review Letters 100~(22) (2008) 228502.

\bibitem{Tse-etal:2010}
C.~K. Tse, J.~Liu, F.~C.~M. Lau, A network perspective of the stock market,
  Journal of Empirical Finance 17~(4) (2010) 659--667.

\bibitem{Gao-Jin:2009}
Z.~Gao, N.~Jin, Complex network from time series based on phase space
  reconstruction, Chaos 19  033137.

\bibitem{Nagy-etal:pigeon_flocks10}
M.~Nagy, Z.~\'{A}kos, D.~Biro, T.~Vicsek, Hierarchical group dynamics in pigeon
  flocks, Nature (London) 464 (2010) 890--894.

\bibitem{Iwayama-etal:2012}
K.~Iwayama, Y.~Hirata, K.~Takahashi, K.~Watanabe, K.~Aihara, H.~Suzuki,
  Characterizing global evolutions of complex systems via intermediate network
  representations, Scientific Reports 2 (2012) 423.

\bibitem{Gao-etal:network2016}
Z.-K. Gao, M.~Small, J.~K. Kurths, Complex network analysis of time series,
  Europhysics Letters 116~(5) (2016) 50001.

\bibitem{Judd-Mees:1995}
K.~Judd, A.~Mees, On selecting models for nonlinear time series, Physica D 82
  (1995) 426--444.

\bibitem{Judd-Mees:1998}
K.~Judd, A.~Mees, Embedding as a modeling problem, Physica D 120 (1998)
  273--286.

\bibitem{Small-Judd:1999}
M.~Small, K.~Judd, Detecting periodicity in experimental data using linear
  modeling techniques, Physical Review E 59 (1999) 1379--1385.

\bibitem{Blinowska:2011}
K.~J. Blinowska, Review of the methods of determination of directed
  connectivity from multichannel data, Medical \& Biological Engineering \&
  Computing 49 (2011) 521--529.

\bibitem{Sameshima-Baccala_book:2014}
K.~Sameshima, L.~A. Baccal\'{a} (Eds.), Methods in Brain Connectivity Inference
  through Multivariate Time Series Analysis (Frontiers in Neuroengineering
  Series), CRC Press, Boca Raton, 2014.

\bibitem{Kaminski-eta:DTF2001}
M.~Kami\'{n}ski, M.~Ding, W.~A. Truccolo, S.~L. Bressler, Evaluating causal
  relations in neural systems: granger causality, directed transfer function
  and statistical assessment of significance., Biological Cybernetics 85 (2001)
  145--157.

\bibitem{Toppi-etal:2012}
J.~Toppi, F.~D.~V. Fallani, G.~Vecchiato, A.~G. Maglione, F.~Cincotti,
  D.~Mattia, S.~Salinari, F.~Babiloni, L.~Astolfi, How the statistical
  validation of functional connectivity patterns can prevent erroneous
  definition of small-world properties of a brain connectivity network,
  Computational and Mathematical Methods in Medicine 2012 (2012) 130985.

\bibitem{Theiler-etal:surrogate92}
J.~Theiler, S.~Eubank, A.~Longtin, B.~Galdrikian, J.~D. Farmer, Testing for
  nonlinearity in time series: the method of surrogate data, Physica D 58
  (1992) 77--84.

\bibitem{galka_book}
A.~Galka, Topics in Nonlinear Time Series Analysis, World Scientific Publishing
  Company, Singapore, 2000.

\bibitem{Michael_book}
M.~Small, Applied Nonlinear Time Series Analysis, World Scientific Publishing
  Company, Singapore, 2005.

\bibitem{Nakamura-etal:SSS_network2016}
T.~Nakamura, T.~Tanizawa, M.~Small, Constructing networks from a dynamical
  system perspective for multivariate nonlinear time series, Physical Review E
  93 (2016) 032323.

\bibitem{Kitagawa-Akaike:modelling78}
G.~Kitagawa, H.~Akaike, A procedure for the modeling of non-stationary time
  series, Annals of the Institute of Statistical Mathematics 30 (1978)
  351--363.

\bibitem{Nakamura-etal:2003}
T.~Nakamura, K.~Judd, A.~Mees, Refinements to model selection for nonlinear
  time series, International Journal of Bifurcation and Chaos 13~(5) (2003)
  1263--1274.

\bibitem{Nakamura-etal:IC2006}
T.~Nakamura, K.~Judd, A.~I. Mees, M.~Small, A comparative study of information
  criteria for model selection, International Journal of Bifurcation and Chaos
  16~(8) (2006) 2153--2175.

\bibitem{Schwarz:SIC}
G.~Schwarz, Estimating the dimension of a model, Annals of Statistics 6 (1978)
  461--464.

\bibitem{Theiler-Prichard:Monte_Carlo_test}
J.~Theiler, D.~Prichard, Constrained-realization monte-carlo method for
  hypothesis testing, Physica D 94 (1996) 221--235.

\bibitem{Zalesky-etal:2012}
A.~Zalesky, A.~Fornito, E.~Bullmore, On the use of correlation as a measure of
  network connectivity, NeuroImage 60 (2012) 2096--2106.

\bibitem{Nakamura-etal:2004}
T.~Nakamura, D.~Kilminster, K.~Judd, A.~Mees, A comparative study of model
  selection methods, International Journal of Bifurcation and Chaos 14~(3)
  (2004) 1129--1146.

\bibitem{Nakamura-Small:2006}
T.~Nakamura, M.~Small, Nonlinear dynamical system identification with dynamic
  noise and observational noise, Physica D 223 (2006) 54--68.

\bibitem{Akaike:AIC74}
H.~Akaike, A new look at the statistical identification model, IEEE
  Transactions on Automatic Control 19 (1974) 716--723.

\bibitem{Kevin:book_chapter2003}
K.~Judd, Building optimal models of time series, in: G.~Gouesbet,
  S.~Meunier-Guttin-Cluzel, O.~M\'enard (Eds.), Chaos and its Reconstruction,
  Nova Science Publications, New York, 2003, pp. 179--214.

\bibitem{Rissanen_book}
J.~Rissanen, Stochastic complexity in statistical inquiry, World Scientific
  Publishing Company, Singapore, 1989.

\bibitem{Rissanen:NML}
J.~Rissanen, Mdl denoising, IEEE Trans on Information Theory 46 (2000)
  2537--2543.

\bibitem{Small-Judd:1998}
M.~Small, K.~Judd, Comparisons of new nonlinear modeling techniques with
  applications to infant respiration, Physica D 117 (1998) 283--298.

\bibitem{Rossler_eqs:1976}
O.~E. R{\"o}ssler, An equation for continuous chaos, Physics Letters A 57~(5)
  (1976) 397--398.

\bibitem{Small-etal:PPS2001}
M.~Small, D.~Yu, R.~G. Harrison, Surrogate test for pseudoperiodic time series
  data, Physical Review Letters 7 (2001) 188101.

\bibitem{Tibshirani:losso1996}
R.~Tibshirani, Regression shrinkage and selection via the lasso, Journal of the
  Royal Statistical Society, Series B 58 (1996) 267--288.

\bibitem{Vidal:SA_book1993}
R.~V.~V. Vidal (Ed.), Applied Simulated Annealing, Vol. 396 of Lecture Notes in
  Economics and Mathematical Systems, Springer-Verlag Berlin Heidelberg,
  Berlin, 1993.

\bibitem{Holland:GA_book1992}
J.~H. Holland, Adaptation in Natural and Artificial Systems: An Introductory
  Analysis with Applications to Biology, Control, and Artificial Intelligence,
  Complex Adaptive Systems, MIT Press, Cambridge, 1992.

\bibitem{Ohlsson:deep_learning_book2011}
S.~Ohlsson, Deep Learning: How the Mind Overrides Experience, Cambridge
  University Press, New York, 2011.

\bibitem{Davis-etal:MVAR_model}
R.~A. Davis, P.~Zang, T.~Zheng, Sparse vector autoregressive modeling, Journal
  of Computational and Graphical Statistics 25 (2016) 1077--1096.

\bibitem{May:logistic_map1976}
R.~M. May, Simple mathematical models with very complicated dynamics, Nature
  (London) 261 (1976) 459--467.

\bibitem{Granger:1969}
C.~W.~J. Granger, Investigating causal relations in by econometric models and
  cross-spectral methods, Econometrica 37~(3) (1969) 424--348.

\bibitem{Jasper:10-20}
H.~H. Jasper, The ten-twenty electrode system of the international federation,
  Electroencephalography and Clinical Neurophysiology 10 (1958) 371--375.

\bibitem{Rapp-etal:2005}
P.~E. Rapp, C.~J. Cellucci, T.~A.~A. Watanabe, A.~M. Albano, Quantitative
  characterization of the complexity of multichannel human eegs, International
  Journal of Bifurcation and Chaos 15~(5) (2005) 1737--1744.

\bibitem{Srinivasan-etal:2007}
R.~Srinivasan, W.~R. Winter, J.~Ding, P.~L. Nunez, Eeg and meg coherence:
  Measures of functional connectivity at distinct spatial scales of neocortical
  dynamics, Journal of Neuroscience Methods 166 (2007) 41--52.

\bibitem{Brunner-etal:2016}
C.~Brunner, M.~Billinger, M.~Seeber, T.~R. Mullen, S.~Makeig, Volume conduction
  influences scalp-based connectivity estimates, Frontiers in computational
  neuroscience 10 (2016) 121.

\bibitem{Steen-etal:2016}
F.~V. de~Steen, L.~Faes, E.~Karahan, J.~Songsiri, P.~A. Valdes-Sosa,
  D.~Marinazzo, Critical comments on eeg sensor space dynamical connectivity
  analysis, Brain Topography (2016) 1--12.

\bibitem{Kaminski-Blinowska:2017}
M.~Kaminski, K.~J. Blinowska, The influence of volume conduction on dtf
  estimate and the problem of its mitigation, Frontiers in computational
  neuroscience 11 (2017) 36.

\bibitem{Haufe-etal:2013}
S.~Haufe, V.~V. Nikulin, K.-R. Muller, G.~Nolte, A critical assessment of
  connectivity measures for eeg data: A simulation study, Neuroimage 64 (2013)
  120--133.

\bibitem{Rapp-etal:1993}
P.~E. Rapp, A.~M. Albano, T.~I. Schmah, L.~A. Farwell, Filtered noise can mimic
  low-dimensional chaotic attractors, Physical Review E 47~(4) (1993)
  2289--2297.

\end{thebibliography}

\end{document}